\documentclass{article}                     % onecolumn (standard format)
\usepackage[T1]{fontenc}
\usepackage[utf8]{inputenc}
\usepackage[english]{babel}
\usepackage{authblk}
\usepackage{graphicx}
\usepackage{afterpage}
\usepackage{array}
\usepackage{booktabs}
\usepackage[inline]{enumitem}
\usepackage{epigraph}
\usepackage{etoolbox}
\usepackage[section]{placeins}
\usepackage{geometry}
\usepackage{hyperref}
\usepackage[sort]{natbib}
\usepackage{pdflscape}
\usepackage{siunitx}
\usepackage[caption=false]{subfig}
\usepackage{tabularx}
\usepackage{threeparttable}
\usepackage[referable]{threeparttablex}
\setcitestyle{aysep={}}
\usepackage[nobiblatex]{xurl}
\hypersetup{
    colorlinks=true,
    linkcolor=blue,
    filecolor=magenta,      
    urlcolor=cyan,
  }

\makeatletter
\patchcmd{\NAT@citex}
  {\@citea\NAT@hyper@{%
     \NAT@nmfmt{\NAT@nm}%
     \hyper@natlinkbreak{\NAT@aysep\NAT@spacechar}{\@citeb\@extra@b@citeb}%
     \NAT@date}}
  {\@citea\NAT@nmfmt{\NAT@nm}%
   \NAT@aysep\NAT@spacechar\NAT@hyper@{\NAT@date}}{}{}
\patchcmd{\NAT@citex}
  {\@citea\NAT@hyper@{%
     \NAT@nmfmt{\NAT@nm}%
     \hyper@natlinkbreak{\NAT@spacechar\NAT@@open\if*#1*\else#1\NAT@spacechar\fi}%
       {\@citeb\@extra@b@citeb}%
     \NAT@date}}
  {\@citea\NAT@nmfmt{\NAT@nm}%
   \NAT@spacechar\NAT@@open\if*#1*\else#1\NAT@spacechar\fi\NAT@hyper@{\NAT@date}}
  {}{}
\makeatother
  
\newcolumntype{Z}{>{\raggedright\let\newline\\\arraybackslash\hspace{0pt}}X}  
\title{Scientometric analysis and knowledge mapping of literature-based
  discovery (1986–2020)\thanks{Authors were supported by the Slovenian Research
    Agency (Grant No. Z5-9352 (AK) and J5-1780 (DH)).}}

\renewcommand{\thefootnote}{\fnsymbol{footnote}}
\author{Andrej Kastrin\thanks{Corresponding author (e-mail: \href{mailto:andrej.kastrin@mf.uni-lj.si}{andrej.kastrin@mf.uni-lj.si})}}
\author{Dimitar Hristovski}
\affil{University of Ljubljana, Faculty of Medicine, Institute for Biostatistics and Medical Informatics, Ljubljana, Slovenia}

\date{\today}
\usepackage[usenames,dvipsnames]{xcolor}
\usepackage{tcolorbox}
\usepackage{lipsum}

\begin{document}


%\titlerunning{Scientometrics of literature-based discovery}        % if too long for running head


%% \institute{A. Kastrin \at
%%               University of Ljubljana, Faculty of Medicine, Institute for
%%               Biostatistics and Medical Informatics, Vrazov trg 2, SI--1000
%%               Ljubljana, Slovenia \\
%%               %Tel.: +123-45-678910\\
%%               %Fax: +123-45-678910\\
%%               \email{andrej.kastrin@mf.uni-lj.si} %  \\
%% %             \emph{Present address:} of F. Author  %  if needed
%%            \and
%%            D. Hristovski \at
%%               University of Ljubljana, Faculty of Medicine, Institute for
%%               Biostatistics and Medical Informatics, Vrazov trg 2, SI--1000
%%               Ljubljana, Slovenia \\
%%               \email{dimitar.hristovski@mf.uni-lj.si}
%% }

%   The correct dates will be entered by the editor
\thispagestyle{empty}
\begin{tcolorbox}[colback=green!5,colframe=green!40!black,title=How to cite this paper?]
This is a pre-print of an article published in Scientometrics. The final authenticated version is available online at \url{https://doi.org/10.1007/s11192-020-03811-z}.
\end{tcolorbox}

\clearpage
\setcounter{page}{1}
\maketitle
\renewcommand{\thefootnote}{\arabic{footnote}}

\begin{abstract}
  Literature-based discovery (LBD) aims to discover valuable latent
  relationships between disparate sets of literatures. This paper presents the
  first inclusive scientometric overview of LBD research. We utilize a
  comprehensive scientometric approach incorporating CiteSpace to systematically
  analyze the literature on LBD from the last four decades (1986--2020). After
  manual cleaning, we have retrieved a total of 409 documents from six
  bibliographic databases and two preprint servers. The 35 years' history of LBD
  could be partitioned into three phases according to the published papers per
  year: incubation (1986--2003), developing (2004--2008), and mature phase
  (2009--2020). The annual production of publications follows Price's law. The
  co-authorship network exhibits many subnetworks, indicating that LBD research
  is composed of many small and medium-sized groups with little collaboration
  among them. Science mapping reveals that mainstream research in LBD has
  shifted from baseline co-occurrence approaches to semantic-based methods at
  the beginning of the new millennium. In the last decade, we can observe the
  leaning of LBD towards modern network science ideas. In an applied sense, the
  LBD is increasingly used in predicting adverse drug reactions and drug
  repurposing. Besides theoretical considerations, the researchers have put a
  lot of effort into the development of Web-based LBD applications. Nowadays,
  LBD is becoming increasingly interdisciplinary and involves methods from
  information science, scientometrics, and machine learning. Unfortunately, LBD
  is mainly limited to the biomedical domain. The cascading citation expansion
  announces deep learning and explainable artificial intelligence as emerging
  topics in LBD. The results indicate that LBD is still growing and evolving.

  %\keywords{Literature-based discovery \and Scientometrics \and Information
  %  visualization \and CiteSpace}
\end{abstract}

\section{Introduction}
\label{intro}

% \setlength{\unitlength}{1pt}
% \epigraphhead[140]{
%\epigraph{\textit{A hidden connection is stronger than an obvious one.}}{---Heraclitus}
% }

Research has shown that scientific output in
terms of original articles, conference proceedings, and books has been
increasing at an accelerated rate~\citep{Bornmann2015}. For instance, the
United States National Library of Medicine adds more than 2000 papers a day to
MEDLINE, the world's leading bibliographic database in the field of life
sciences. Faced with information overload, scientists often miss valuable
pieces of knowledge relevant to their research interests.

Given the massive amounts of scientific data generated every day, extracting
and pinpointing relevant information becomes an important pursuit, albeit a
challenging one. It is a challenging task to join disparate scientific pieces
of information into a comprehensive body of knowledge. Nowadays, computational
methods are used to complement manual knowledge discovery from textual
data. Literature-based discovery (LBD) is an interesting yet highly
challenging research paradigm that uses computational algorithms for mining
scientific literature. In modern text mining, LBD research plays an important
role. LBD has been successfully utilized in various application areas
including life sciences~\citep{Pyysalo2019}, humanities~\citep{Cory1997}, and
counterterrorism~\citep{Jha2016}.

By definition, LBD is a text mining approach for automatically generating
research hypotheses~\citep{Smalheiser2017}. The main aim of LBD is to
stimulate and support human creativity to find important connections between
disparate literatures by identifying hidden, previously unknown relationships
from existing knowledge. The LBD approach was initiated by
\citet{Swanson1986a}, who discovered that dietary fish oil might be used to
treat Raynaud's disease. This discovery was based on the observation that
Raynaud's disease lowers blood viscosity, reduces platelet aggregation and
inhibits vascular reactivity. (Raynaud's disease exhibits excessively reduced
blood flow in response to cold or emotional stress, causing discoloration of
the fingers, toes, and occasionally other areas.) Swanson's hypothesis was
later verified in vivo by~\citet{DiGiacomo1989}. Nowadays, LBD is an
interdisciplinary research field and it is considered as a branch of both
computer science and information science.

\begin{figure}[htb]
\centering
\includegraphics[scale=1.2]{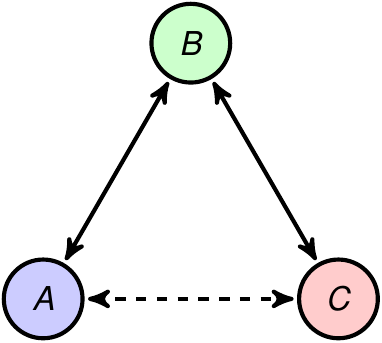} 
\caption{Swanson's ABC discovery model. The model contains three concepts:
  start ($A$), intermediate ($B$), and target ($C$). The LBD process begins
  with retrieving $AB$ and $BC$ relationships. Next, we combine associations
  with the same intermediates. Finally, we get a list of $AC$
  relationships. If there is no prior mention of a particular $AC$ connection,
  we formulate a hypothesis of a potential novel relationship between $A$ and
  $C$ concepts.}
\label{fig:abc_model}
\end{figure}

Swanson's pioneering methodology is based on the presumption that there exist
multiple complementary and non-intersecting knowledge domains in the
scientific literature~\citep{Swanson1986b}. Knowledge in a given domain may be
related to knowledge in another domain, but without the relationship being
known. The Swanson's LBD paradigm relies on the notion of concepts relevant to
three literature domains: $A$, $B$, and $C$ (Figure~\ref{fig:abc_model}). For
instance, let us suppose we have found a link between a disease $A$ and a gene
$B$. Next, suppose that another research group has published the effect of a
drug $C$ on the gene $B$. The use of LBD methodology may propose an $AC$
relation, suggesting that the drug $C$ may potentially treat the disease
$A$. Such a latent link may represent a novel hypothesis for a potential, yet
unconfirmed relationship.

\begin{figure}[htb]
\centering
\subfloat[]{\includegraphics[scale=1.2]{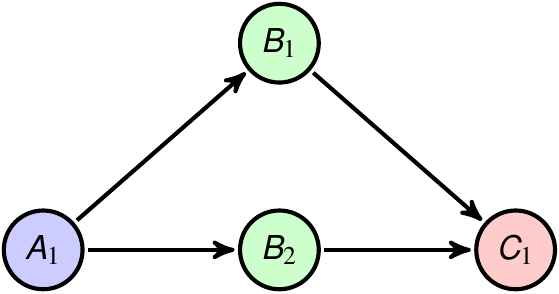}\label{fig:closed_lbd}}
\hspace{5mm}
\subfloat[]{\includegraphics[scale=1.2]{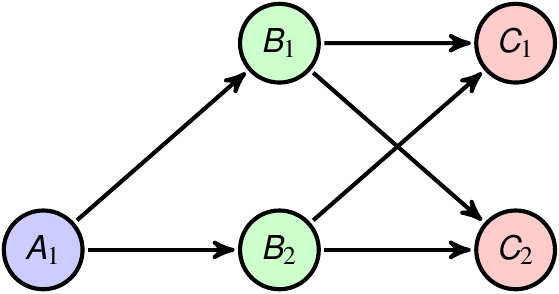}\label{fig:open_lbd}}
\caption{Closed \protect\subref{fig:closed_lbd} and open
  \protect\subref{fig:open_lbd} discovery model. In closed discovery, we seek
  for intermediate concepts ($B_1$, $B_2$, \ldots, $B_n$) that connect a start
  concept ($A$) and a target concept ($C$). In open discovery mode, only the
  start concept ($A$) is given and the goal is to identify target concepts
  ($C_1$, $C_2$, \ldots, $C_m$) through intermediate concepts ($B_1$, $B_2$,
  \ldots, $B_n$).}
\label{fig:lbd_modes}
\end{figure}

The ABC model could be used in two ways; open discovery and closed
discovery~\citep{Weeber2001}. The former is typically used as a hypothesis
generation process, and the latter as a hypothesis testing process. In the
open discovery (Figure~\ref{fig:closed_lbd}), we start with a concept $A$
(e.g., disease) and try to find intermediate concepts $B$ (e.g., molecular
functions) that play a role in explaining the concept $A$. In the second step,
we need to identify concepts $C$ (e.g., genes) that are directly connected to
concepts $B$. Finally, we hypothesize that the concept $C$ is related to the
concept $A$ through the intermediate $B$. On the other hand, a closed
discovery (Figure~\ref{fig:open_lbd}) starts with concepts $A$ and $C$ and
tries to find intermediate $B$s. The more intermediates we find, the more
plausible is the tested hypothesis. Although simple, Swanson's ABC model is
widely accepted in the LBD community. However, in the last decade, researchers
have proposed several other discovery strategies, including discovery
browsing~\citep{Wilkowski2011}, outlier detection~\citep{Petric2012},
entitymetrics~\citep{Ding2013}, link prediction~\citep{Kastrin2016},
analogical reasoning~\citep{Mower2017}, heterogeneous bibliographic
networks~\citep{Sebastian2017b}, and neural networks~\citep{Crichton2018}. All
these approaches persuasively improve the performance of the basic ABC
model~\citep{Smalheiser2017}.

LBD has found greatest utility in the biomedical domain. For instance, the LBD
methodology has been applied to identify disease candidate genes for
polymicrogyria~\citep{Hristovski2005} or to propose potential treatments for
Parkinson's disease~\citep{Kostoff2008}. LBD is increasingly used for drug
repurposing~\citep{Yang2017} and for better understanding and prediction of
adverse drug events~\citep{Shang2014}. Last but not least, LBD has been
applied in a framework for cross-domain recommendation for biomedical research
collaboration~\citep{Hristovski2015}. The lack of comprehensive ontologies and
tools (e.g., UMLS~\citep{Bodenreider2004}, SemRep~\citep{Rindflesch2003},
SemMedDB~\citep{Kilicoglu2012}) is the main reason why LBD has not been widely
adopted outside the biomedical domain~\citep{Hui2019}.

Scientometric analysis has a critical role in strategic science planning,
policy, and research performance evaluation. Scientometrics concerns a broad
range of research methodologies and technologies including modern statistical
analysis and visualization. The primary goal of the scientometric analysis is
to assess the performance of a research unit of interest (e.g., scholar,
journal, institution, discipline, country) and examine and summarize its
knowledge structure and evolution. In recent decades, a scientometric review
has been broadly adopted to quantitatively evaluate the previous research
activities, track the transformative processes, understand knowledge
landscapes, and predict emerging trends in various scientific
fields~\citep{Chen2017}. The roots of scientometrics could be traced back to
the early 20th century~\citep{Godin2006}, however, the main methodological
tools were developed in the 1960s \citep{Price1965,Pritchard1969}.

Contemporary scientometrics incorporates two different but methodologically
complementary research approaches~\citep{Noyons1999}:
\begin{enumerate*}[label=(\roman*)]
\item performance analysis and
\item science mapping.
\end{enumerate*}
The former procedure includes, for instance, various
counts (e.g., publications and citations by authors, countries, and
institutions), burst detection~\citep{Kleinberg2003}, or $h$-index
analysis~\citep{Hirsch2005}. On the other hand, science mapping employs
different spatial and temporal representation techniques to examine the
structural and dynamic properties of scientific
research~\citep{Chen2017, Chen2013}. Frequently used tools are co-citation
analysis~\citep{Small1973} and co-word analysis~\citep{Callon1983}. The
development of scientometric software goes hand-in-hand with the advancement
in information sciences and novel visualization
approaches~\citep{Chen2017, Chen2009b}. The most popular software tools for
bibliographic analysis are, among others, CiteSpace~\citep{Chen2006},
VOSviewer~\citep{vanEck2009}, SciMAT~\citep{Cobo2012}, and bibliometrix
package for \textsf{R}~\citep{Aria2017}.

To the best of our knowledge, there is no detailed scientometric-based
scientific review of the LBD research field currently, although---as we will
see in the Related work section---at least two papers try to fill this
gap~\citep{Thilakaratne2019a, Chen2019}. However, the growth in LBD literature
necessitates a detailed scientometric review. This study aims to extend
previous traditional reviews of the LBD literature \citep{Smalheiser2017,
  Thilakaratne2019a, Davies1989, Weeber2005, Bekhuis2006, Smalheiser2012,
  Ahmed2016, Sebastian2017a, Henry2017, Gopalakrishnan2019,
  Thilakaratne2019b}; we conduct a quantitative scientometric analysis on
publications retrieved from Web of Science (WoS), Scopus, PubMed, and other relevant
bibliographic databases since the inception of LBD in 1986. We analyze
bibliographic metadata from citation indexes (i.e., titles, abstracts, journal
names, author names, author addresses) to infer production, impact, fields of
interests, and general characteristics of the LBD literature and create a
scientometric profile of LBD research. At the same time, we try to interpret
the findings from the perspective of LBD experts. Specifically, to address the
LBD field, this paper
\begin{enumerate*}[label=(\roman*)]
\item provides a comprehensive overview of the research evidence using the
  scientometric analysis by summarizing the majority of the papers published
  in the last 35 years;
\item identifies key authors, countries, institutions,
  and main describable keywords related to the research area;
\item deduces the most noticeable end emerging research themes in the field of
  LBD; and
\item compares the most influential works based on citation statistics.
\end{enumerate*}
The findings of this study could be relevant to different
stakeholders. Particularly, the presented analysis may be relevant to
researchers new to LBD to orient in the field, to identify knowledge gaps, and
to move the LBD field forward.

\section{Related work}

LBD is a complex, continually evolving, and collaborative research field. To
the best of our knowledge, at least ten traditional literature reviews have
been published to elucidate the extent of knowledge in the LBD research
domain. Below we give a brief description of each of them.

Only three years after Swanson's first paper on LBD, \citet{Davies1989}
published an interesting theoretical paper on the creation of new knowledge by
information retrieval. This article provides an in-depth review of previous
work on generating knowledge through information retrieval and presents
methods to retrieve latent knowledge from the literature (e.g., serendipity
browsing, proper search strategies, relational indexing, and artificial
intelligence). Together with some recent theoretical
papers~\citep{Chen2009b,Uzzi2013}, Davies's paper is a must-read for all
researchers who seek to deeply understand the formalistic foundations of LBD.

With the rise of various technologies (e.g., JavaScript) in the mid-1990s that
stimulated the development of interactive Web applications, the LBD community
started to build online LBD tools and services. \citet{Weeber2005} provided a
review of methodology and LBD tools that had been developed until 2005. The
authors described Arrowsmith~\citep{Swanson1997},
BITOLA~\citep{Hristovski2005}, Manjal~\citep{Srinivasan2004},
LitLinker~\citep{Pratt2003}, ACS~\citep{Eijk2004},
IRIDESCENT~\citep{Wren2004a}, and Telemakus~\citep{Fuller2004}. To our
knowledge, only Arrowsmith and BITOLA are still available from that list to
the broad research public.

\citet{Bekhuis2006} described LBD in the context of conceptual biology and the
broader domain of text mining. The author provided a general background for
knowledge discovery, a brief review of Swanson's ideas, and a short discussion
of approaches for hypothesis discovery. Her review is complementary to the
overviews published by \citet{Cohen2005} and \citet{Jensen2006} around the
same time.

As Swanson's collaborator, Smalheiser wrote two review papers on LBD
\citep{Smalheiser2017,Smalheiser2012}. In the first article Smalheiser
provided a critical overview of the then prevalent ABC model. He concluded
that the ABC paradigm was not wrong, however, it was only one of the many
approaches to LBD that could stimulate the development of a new generation of
LBD tools. Moreover, he advocated that we urgently needed some sort of
objective function (i.e., interestingness measures) for filtering out the
output of LBD systems. Smalheiser's second review was written from a more
personal perspective. The author discussed Swanson's contributions to LBD and
gave an outline of its future directions.

\citet{Ahmed2016} provided the first attempt to systematically classify LBD
methods and approaches. The author defined LBD exclusively through the lens of
information retrieval. The paper identified three approaches that were most
often used as a basis for LBD: vector space model, probabilistic methods, and
inference network.

Nearly in the same year, \citet{Sebastian2017a} and \citet{Henry2017},
published extensive papers on LBD. The first group of authors provided an
in-depth discussion on a broad palette of existing LBD approaches and offered
performance evaluations on some recent emerging LBD methodologies. The latter
authors likewise introduced historical and modern LBD approaches and provided
an overview of evaluation methodologies and current trends. Both papers
provided a general unifying framework for the LBD paradigms, its
methodologies, and tools. The next review was published recently by
\citep{Gopalakrishnan2019}. Their paper provided a comprehensive analysis of
the LBD field, and the paper served as a methodological introduction behind
particular tools and techniques. The authors provided a detailed discussion of
the key LBD systems through the critical analysis of selected influential
papers. They also summarized recent research trends and identified future
directions of LBD. \citet{Thilakaratne2019a} analyzed the methodologies used
in LBD using a novel classification scheme (i.e., the main points of the
review were computational techniques, central research topics, available
tools, and applications) and provided a timeline with key milestones in LBD
research. The authors also identified the current trends in LBD regarding
publication over years, top cited papers, and top authors. However, they
considered only journals and conference proceedings for the review. In their
second paper, \citet{Thilakaratne2019b} presented a large-scale systematic
review of the LBD workflow by manually analyzing 176 LBD papers. Although
these reviews successfully provide qualitative insight into the field of LBD
through dissecting the research evidence and appropriate classification of
research themes, their analysis was manual and did not offer quantitative
examination. Recently, \citet{Chen2019} performed the first knowledge mapping
of the LBD field. They use the LBD domain as a proxy to illustrate an
intuitive method to compare multiple search strategies in order to identify
the most representative body of scientific publications.

Although the aforementioned reviews are quite recent, an in-depth quantitative
analysis in the LBD research field is urgently needed, to provide newcomers,
researchers, and also clinicians with a state-of-the-art scientometric
overview of the field. It is also important to keep researchers informed about
emerging trends and essential turning points in the expansion of domain
knowledge.

\section{Methods}

In this section, we first outline the data collection procedure. Then we
proceed with computational methodology and techniques applied in scientometric
analysis, and finally, we conclude with a description of tools that have been
used.

\subsection{Bibliographic data collection}

We used the most authoritative bibliographic databases as the data sources for
retrieving publications and related metadata in the LBD research domain,
including WoS (\url{https://clarivate.com/products/web-of-science}; Clarivate
Analytics, Philadelphia, Pennsylvania, USA), Scopus
(\url{https://www.scopus.com}; Elsevier, Amsterdam, Netherlands), PubMed
(\url{https://www.ncbi.nlm.nih.gov/pubmed}; National Library of Medicine,
Be\-thesda, Maryland, USA), ACM Digital Library (\url{https://dl.acm.org};
Association for Computing Machinery, New York, New York, USA), IEEE Xplore
(\url{https://ieeexplore.ieee.org}; Institute of Electrical and Electronics
Engineers, Piscataway, New Jersey, USA), and SpringerLink
(\url{https://rd.springer.com}; Berlin, Germany). In addition, we used two
preprint servers, namely arXiv (\url{https://arxiv.org}) and bioRxiv
(\url{https://www.biorxiv.org}). WoS and Scopus both offer more or less
comprehensive synopsis of the world’s research evidence in science,
technology, medicine, social science, and arts and humanities. Scopus indexes
literature dating back to 1970, while WoS covers even older publications as
its index goes back to 1900. Preliminary analysis and our own experiences
indicated that the prevailing body of LBD literature originates from
biomedicine. To this end, we also included PubMed which is a primary
bibliographic database in the field of biomedicine. ACM Digital Library and
IEEE Xplore was used predominantly for retrieving conference proceedings. Due
to the fact that more and more authors publish their manuscripts on preprint
servers, we also included arXiv and biorXiv. The former is an open-access
repository of preprints for natural sciences and the latter for life
sciences. Finally, to reduce the risk of losing important documents, we also
collected all relevant references from recent LBD reviews
\citep{Smalheiser2012, Henry2017, Sebastian2017a, Smalheiser2017,
  Gopalakrishnan2019, Thilakaratne2019a, Thilakaratne2019b}.

Our objective was to include a complete universum of publications on LBD. For
this purpose, we designed a search strategy to identify records where LBD
related terms were mentioned in the title, abstract, or among keywords of the
bibliographic citations. The detailed search strategy for each database is
shown in Table~\ref{tab:queries}. The time span was set between the years 1986
and 2020, since Swanson's first paper until now. We applied no language,
geographic, or any other constraints on the database retrieval procedure. Each
bibliographic record consists of metadata about the publication, including a
list of authors, title, abstract, author keywords, author affiliation, as well
as number of citations, and a list of references cited by the publication. We
included publications of all source types including journals, conference
proceedings, book series, and books. We included the following document types:
article, article in press, review, letter, editorial, note, short survey,
conference paper, book, book chapter, erratum, and conference review. For some
bibliographic records a manual inspection of the underlying paper was needed
to identify missing bibliographic details (e.g., author's affiliation). Author
names normalization and disambiguation was not necessary. We have collected
and downloaded all full-texts in PDF format and imported them into the Zotero
reference manager. We removed all duplicate records using the Jaro-Winkler
distance between pairs of titles as implemented in the \textsf{stringdist}
package in \textsf{R}.

The bibliographic records that satisfy the search strategy were included for
further analysis. The first author (AK) conducted a manual verification to
ensure that each publication was closely related to the LBD field. During this
first check, based on screening titles, abstracts, and keywords of the
publications, AK eliminated the irrelevant publications. In the second step,
both authors have evaluated the remaining publications (the relevant ones). We
discussed all discrepancies until consensus has been reached. If necessary, one
of us read the full paper to understand the content and the background of the
paper and decided whether to include it in the analysis. The detailed review
framework is depicted in the Results section in Figure~\ref{fig:prisma}.

\subsection{Data analysis}

The bibliographic records from different databases were first merged into the
core dataset. We have paid special attention to cross-checking to ensure
consistency of the data. First, we identified the annual production of LBD
literature. We statistically described the annual distribution of publications
by Price's law~\citep{Price1963}, which postulates an exponential growth of
scientific production in a given domain over a predefined survey
period. Hence, we first plot publication frequency against year, and then we
apply the best-fitting linear and exponential functions to the data. If the
latter has a better fit then the former, than we can consider the distribution
as fulfilling the Price's law.

Next, we prepared and summarised the statistics on most prolific authors,
countries, institutions, and journals. We identified the first author's
affiliation, corresponding institution name, and country from the author
address information. In cases where the author address was missing, we
identified proper affiliation using the ``Author Search'' function in WoS or
Scopus. The impact factor of the journals was determined as a five-year impact
factor from Journal Citation Reports (\url{https://jcr.clarivate.com};
Clarivate Analytics, Philadelphia, PA, USA). We identified the most prolific
institutions based on the affiliation of the first author of a given
publication. We evaluated the distribution of publications among journals
regarding whether they followed Bradford's law~\citep{Bradford1934}, which
states that if we sort journals by the number of articles published and then
assign them to three groups, with each group publishing one-third of all
articles, then one should identify the number of journals in each zone as the
ratio $1 \colon n \colon n^2$, where $n$ is defined as the Bradford
multiplier. In other words, a few core journals account for one-third of all
papers published within a body of investigated literature, whereas many other
journals publish only a few papers.

It is known from the early days of Gestalt psychology that the whole is
usually more than the pure sum of its parts. In addition to studying
individual scholars, it is important to study their interrelations, and how
such relations evolve in time, respond to internal events and external
perturbations~\citep{Chen2013}. Hence we covered the research characteristics
both at the entity level (i.e., individual researcher, organization, or
country) and at the complex network level. Besides simple counting, we used
two scientometric techniques to elucidate the relationship structure and
dynamics of LBD research:
\begin{enumerate*}[label=(\roman*)]
\item co-authorship analysis (COA) that seeks author
co-occurrences, and
\item document co-citation analysis (DCA) that tries to
summarize the citation structure and provide a glimpse of the relations
between papers.
\end{enumerate*}

Empirical evidence shows that DCA can successfully reveal the latent scientific
structure of an investigated research domain~\citep{Small1973}. Each scientific
paper usually cites a number of other articles. In DCA we represent these
references as nodes and the links between the nodes represent how often a pair
of references are cited together. The underlying assumption of DCA is that the
references are contextually related if they are frequently cited
together~\citep{Chen2010}. CiteSpace offers several selection criteria for nodes
to control the size of the network (e.g., $g$-index, Top N, and threshold
interpolation). Top N is the simplest approach where the top $N$ nodes are
selected with the highest frequency in each time slice. On the other hand, the
$g$-index reflects the global citation performance of a set of
articles~\citep{Egghe2006}. However, we empirically choose the Top N measure. We
have tried several configurations with different $k$ and $N$ values for
$g$-index and Top N, respectively. Choosing Top N selection with $N = 50$
performs best in identifying meaningful and interpretable clusters. To
facilitate the interpretation of the DCA network, we performed cluster analysis
which partitions the co-citation network into non-overlapping clusters. Each
cluster is characterized by the references that are tightly interconnected
within a cluster and exhibit weak connections among different clusters. We
measured the quality of the partitioning process using the cluster's silhouette
width, where greater width reflects higher homogeneity of the cluster.

In the next step, we extended the basic DCA approach with Cascading Citation
Expansion (CCE). CCE is a sophisticated approach that can be used to overcome
the limitations of keyword-based queries and to optimize the quality of a data
collection step in systematic reviews~\citep{Chen2019}. CCE has its roots in the
citation indexing proposed by~\citet{Garfield1955}. Two papers are likely to be
important for a particular research field if one cites the other, even though
they do not have common keywords. Citation expansion is a process of finding
relevant papers based on a list of citation links. The expansion can be done in
two directions. The forward expansion will find citing articles for a particular
paper, while the backward expansion will find papers that are on a reference
list of that paper~\citep{Chen2018}. In this paper, we used a 2-generation
forward expansion process\footnote{A 2-generation forward expansion collects all
  papers connecting to the seed paper with two-step citation paths.} and
utilized a procedure as implemented in CiteSpace with direct access to the
Dimensions API (\url{https://www.dimensions.ai}; Digital Science \& Research
Solutions, London, England). For more details on CCE, we refer the reader to
\citet{Chen2018} and \citet{Chen2019}.

The nodes in a network play various roles. For instance, a node may be central
in a localized region of nodes (i.e., hub) or act as a connector between
disparate clusters of nodes (i.e., broker). The importance of a particular
node in a network is measured by various centrality measures. In this regard,
we measured two types of centrality:
\begin{enumerate*}[label=(\roman*)]
\item betweenness, which identifies nodes that are tightly connected to each
  other in terms of hubs; and
\item brokerage, which determines nodes that filter, control, and alter the
  flow of information among different groups of nodes.
\end{enumerate*}
The procedure for computing betweenness centrality is already implemented in
CiteSpace. To compute brokerage, we first exported the desired network to
\textsf{R} and then used the \texttt{brokerage()} function from the
\textsf{sna} package to perform the brokerage analysis of \citet{Gould1989}.

Next, we detected the burst strength of the authors, institutions, journals,
countries, and keywords. The burst strength characterizes how great the change
is in the item's frequency that triggered the burst. In this study, we used
the original Kleinberg's~\citeyearpar{Kleinberg2003} burst detection algorithm,
which can identify sudden increases in frequency over time. In the paper, we
reported only statistically significant bursts, together with the burst start
and end.

\subsection{Software}

Data preprocessing was performed using custom Bash and Python scripts. The
main part of the data analysis and visualizations was performed in CiteSpace
(ver. 5.6.R5)~\citep{Chen2006} and \textsf{R} using the \textsf{bibliometrix}
package~\citep{Aria2017}. We decided to use this package because it greatly
facilitates reproducible analysis, although many other excellent software
packages for scientometric analysis exist in the community (e.g.,
VOSviewer~\citep{vanEck2009} or SciMAT~\citep{Cobo2012}). The programming
scripts to reproduce the results of our analysis are freely available in the
GitHub repository \url{https://github.com/akastrin/lbd-review}. A
comprehensive data archive is available at Zenodo
(\url{https://doi.org/10.5281/zenodo.3884422}) and includes tabulated
bibliographic data.

\section{Results}

In this section, we present the results of the performed analysis. First, we
analyze the authors' contributions to the body of LBD and provide performance
statistics for the LBD production across countries and institutions. In the
second part of this section, we delve into science mapping and try to
understand the intellectual base of LBD research through the analysis of
keywords and co-citation patterns.

\begin{figure}[htb]
\centering
\includegraphics[scale=0.9]{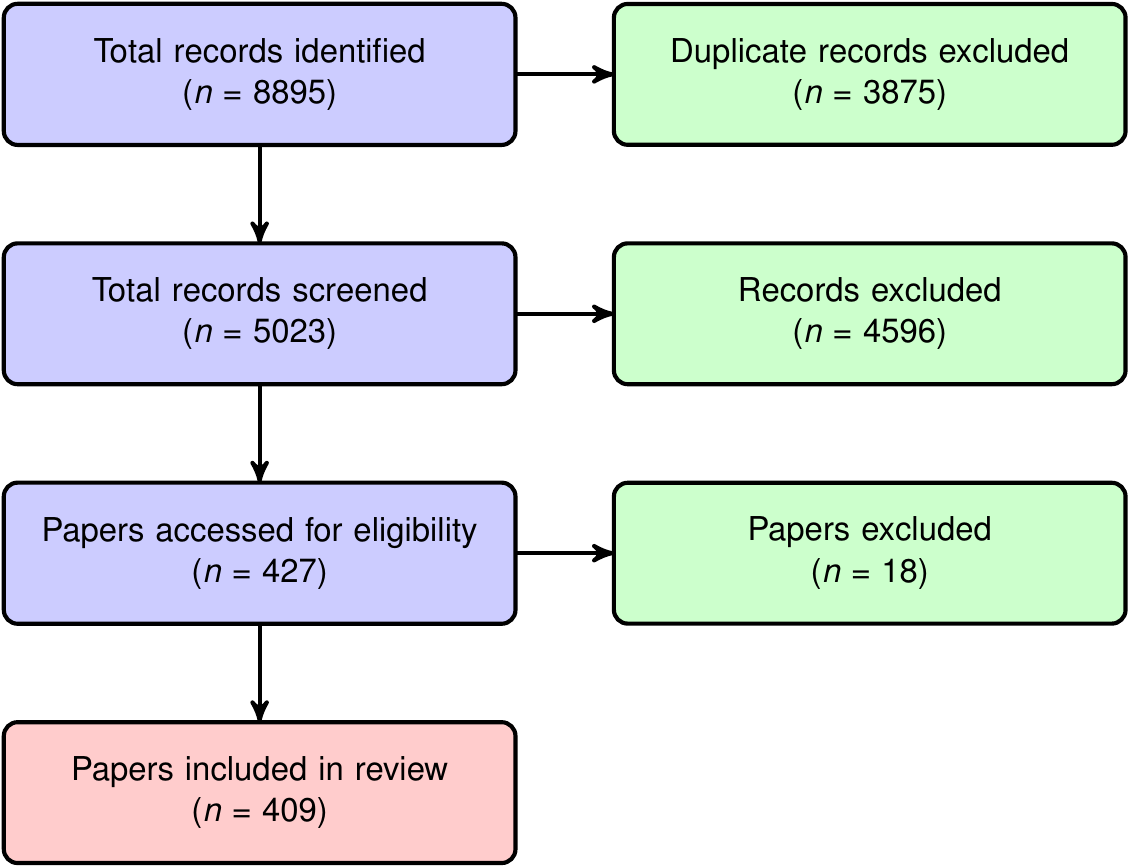}
\caption{PRISMA diagram}
\label{fig:prisma}
\end{figure}

In Figure~\ref{fig:prisma} we depict a four-phase flowchart based on the
PRISMA recommendations~\citep{Liberati2009}. Using the search strategy
described previously in the Methods section, we first retrieve $n = 8895$
bibliographic records. Next, we remove duplicate publications ($n = 3875$)
after which $n = 5023$ records remain. We manually screen the titles and
abstracts and exclude $n = 4596$ records that are not relevant to LBD. The
second screening is performed on full-text publications; in this phase, we
exclude additional 18 non-relevant publications. Finally, 409 publications
remain for further analysis.

\subsection{Retrieved literature}

The number of publications indexed by different bibliographic databases (e.g.,
WoS, Scopus, and PubMed) on the same research subject tends to
differ. Therefore, to extract a more reliable and valid set of data, we use
eight databases. We include publications published between 1986 and 2020 that
represent the complete active period of publication in LBD. A search of the
databases was performed on 1st April 2020. Consequently, publications that
were indexed after this date might not have been captured in our study. The
final set of bibliographic records covers a time period of 35 years
(1986--2020) beginning with Swanson's first paper on the
LBD~\citep{Swanson1986a}. The majority of the records are original articles
($n = 236$), followed by conference papers ($n = 127$), review papers
($n = 24$), book chapters ($n = 11$), and other material ($n = 11$; i.e., book
review, letters to the editor, and reports). All documents were published in
$n = 224$ different sources. As of April 1, 2020, the complete set of
publications had been cited $n = \num{10198}$ times.

\begin{table}[htb]
\centering
\caption{Queries used and statistics of the document retrieval process}
\begin{threeparttable}
\begin{tabular}{lS[table-format=4.0]S[table-format=4.0]S[table-format=3.0]S[table-format=3.0]S[table-format=3.0]S[table-format=3.0]S[table-format=2.0]S[table-format=3.0]S[table-format=4.0]}
\toprule
Query & {WoS} & {Scopus} & {PubMed} & {ACM} & {IEEE} & {Springer} & {ArXiv} &
                                                                              {biorXiv}
  & {Total} \\
\midrule
Query 1\tnote{a} &  203	&  254 &  92 &  39 &  22 & 195 &  6 &  20 &  831 \\
Query 2\tnote{b} &   14	&   23 & 109 &  10 &   3 &   0 &  2 &   0 &  161 \\
Query 3\tnote{c} &   19	&   16 &   0 &   2 &   0 &   0 &  0 &   0 &   37 \\
Query 4\tnote{d} & 1531	& 2135 & 990 & 599 & 336 & 689 & 47 & 529 & 6856 \\
Query 5\tnote{d} &   46	&   25 &   0 &  17 &   1 &  67 &  0 &  11 &  167 \\
\midrule  
\multicolumn{9}{l}{\citet{Smalheiser2012}} & 53 \\
\multicolumn{9}{l}{\citet{Henry2017}} & 96 \\
\multicolumn{9}{l}{\citet{Smalheiser2017}} & 65 \\
\multicolumn{9}{l}{\citet{Sebastian2017a}} & 138 \\
\multicolumn{9}{l}{\citet{Gopalakrishnan2019}} & 129 \\
\multicolumn{9}{l}{\citet{Thilakaratne2019a}} & 224 \\
\multicolumn{9}{l}{\citet{Thilakaratne2019b}} & 138 \\
\midrule
\multicolumn{9}{l}{Total} & 8895 \\
\bottomrule
\end{tabular}
\begin{tablenotes}[flushleft,para]\footnotesize
\item[a] \texttt{``literature based discovery'' OR ``literature based discoveries''} \\
\item[b] \texttt{``literature based knowledge discovery'' OR ``literature based
  knowledge discoveries''} \\
\item[c] \texttt{``literature related discovery'' OR ``literature related
  discoveries''} \\
\item[d] \texttt{``hypothesis generation'' OR ``hypotheses generation''} \\
\item[e] \texttt{``undiscovered public knowledge''} \\
\end{tablenotes}
\end{threeparttable}
\label{tab:queries}
\end{table}

\subsection{Performance bibliometric analysis}

\subsubsection{Publication evolution over the years}

\begin{figure}[htb]
\centering
\includegraphics[width=.95\columnwidth]{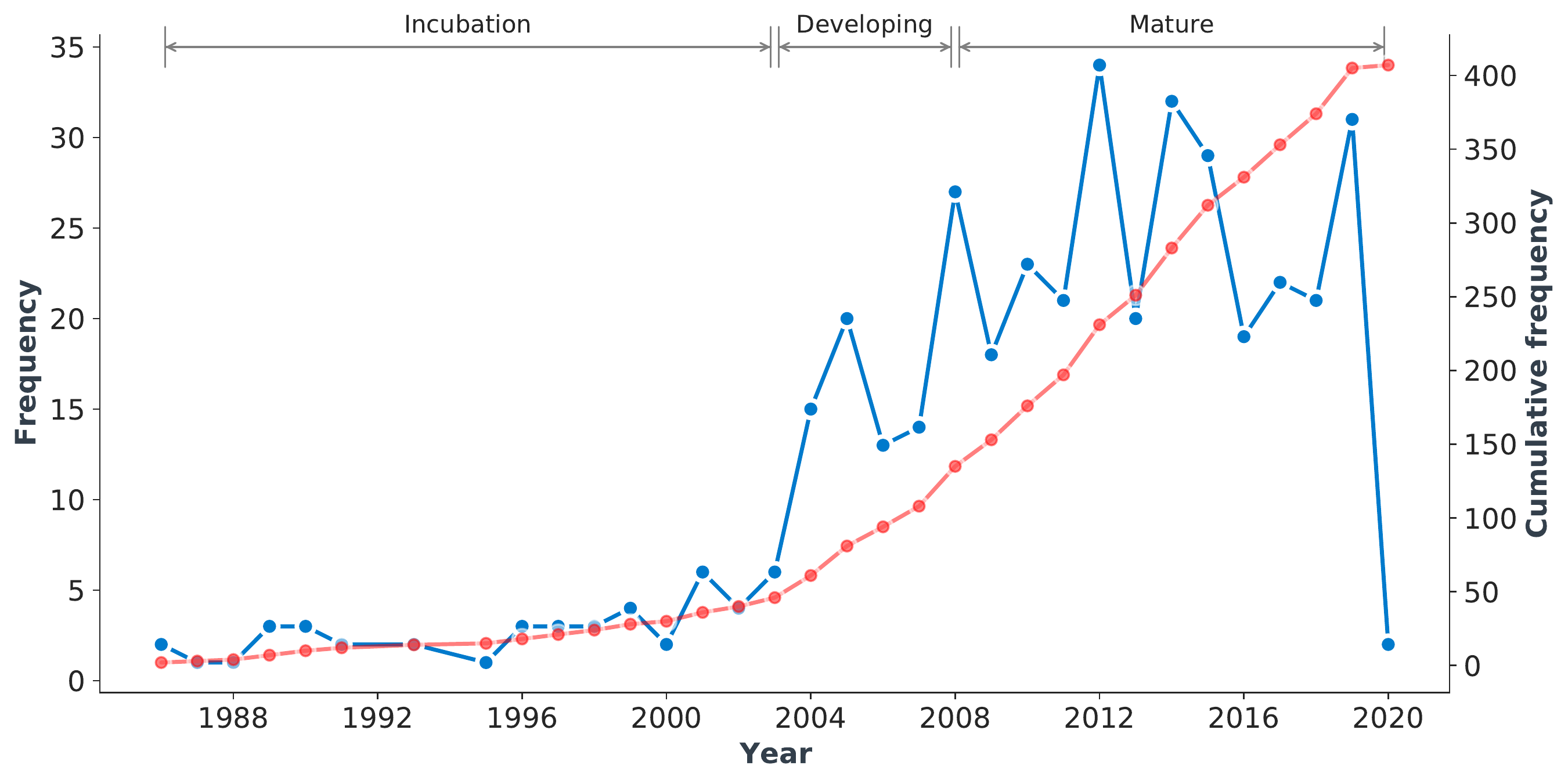}
\caption{A number of documents published annually in the LBD domain
  (1986--2020). Red dashed line represents cumulative frequency.}
\label{fig:annual_distribution}
\end{figure}

The analysis of publication behavior over time might demonstrate the
developmental trend from the macroscopic perspective. The maximum number of
papers ($n = 34$) were published in 2012. It is noteworthy that the term
``Literature Based Discovery'' was included in Medical Subject Headings (MeSH)
vocabulary in 2013, indicating its high bibliographic
importance. Figure~\ref{fig:annual_distribution} depicts the changing pattern
of publications in our data set from 1986 until 2020. A reader can observe
that the number of publications on LBD increased slowly from 1986 to
mid-2000s, but since then it has been increasing significantly. This fact
indicates that the field of LBD has acquired significant attention in the last
decade.

The 35 years' time span could be naturally clustered into three phases
according to the published papers per year:
\begin{enumerate}
\item Incubation phase (1986--2003). In this phase, the number of publications
  was small, and the growing trend was more or less low. In the mid-1980s
  \citet{Swanson1986a} published the seminal paper on LBD. In this period,
  first simple experiments for automating LBD were
  performed~\citep{Gordon1996, Lindsay1999} and the basic terminology was
  refined~\citep{Weeber2001}.
\item Developing phase (2004--2008). In this phase the empirical evidence
  gradually increased, however, the number of publications per year, with the
  exception of the year 2005, was still less than 20. In this phase, the
  foundational aspects of automated LBD were solidified and prepared the basis
  for more advanced investigations in the future. Also, the first book
  dedicated solely to LBD was published~\citep{Bruza2008}.
\item Mature phase (2009--2020). The number of papers published in this period
  is significantly higher in comparison to the previous two phases. Research
  has entered a peak period and even demonstrated a booming tendency. In this
  period researchers developed a plethora of new methods and techniques for
  LBD. Additionally, the first two extensive review papers were
  published~\citep{Sebastian2017a, Henry2017}.
\end{enumerate}

We fit linear and exponential regression models to test whether the annual
distribution of publications followed Price's law. Both models were
statistically significant ($p < 0.05$). The linear model achieved a
coefficient of determination of $R^2 = 0.61$, while the exponential model
fitted better with a coefficient of determination of $R^2 = 0.65$. Therefore,
we could conclude that the annual production of publications follows Price's
law. The annual growth rate was \SI{8.66}{\percent}, where we omit the last
year from the calculation.

\subsubsection{Authors}

The authors are the driving force in research. It is the task of every
scientist to make a meaningful contribution to the body of knowledge and thus
(co-)shape the development of the field in which (s)he works.

Our analysis identifies 802 distinct authors. The majority of the authors
write in collaboration with colleagues ($n = 766$). On average, we have
detected 3.68 ($\textit{SD} = 2.73$) authors per document and 1.90
($\textit{SD} = 2.57$) documents per author. The authors with the highest
number of publications and citations have a tendency to be the scientists who
drive the research field and have a casting vote for its development. The 10
most prolific authors are presented in Table~\ref{tab:top_authors}. Rindflesch
clearly holds the first position with 37 publications, although he is the
first author in only one LBD paper. As stated above, Lotka's law states that a
small portion of researchers is responsible for most of the publications,
whereas the majority contribute a very small number of papers. The discrepancy
between the observed values and the expected frequencies according to Lotka’s
law has been evaluated using the nonparametric Kolmogorov-Smirnov (KS)
goodness-of-fit test. KS test reveals no statistically significant differences
between the observed and the actual publication numbers ($D = 0.28$,
$p = 0.491$).

\begin{table}[h]
\centering
\caption{Top 10 authors based on the total number of publications}
\begin{threeparttable}
\begin{tabularx}{\columnwidth}{S[table-format=2.0]lZS[table-format=2.0]S[table-format=4.0]S[table-format=2.0]S[table-format=2.0]S[table-format=1.2]}
\toprule    
{Rank} & Author	& Institution & {\textit{NP}} & {\textit{TC}} & {$h$} & {$h_s$} & {$b_i$} \\
\midrule
1 & Rindflesch TC & National Library of Medicine, USA & 37 & 1617 & 24 & 16 &  0.04 \\
2 & Kostoff RN & Georgia Institute of Technology, USA & 23 & 3016 & 29 & 15 & 0.00 \\
3 & Hristovski D & University of Ljubljana, Slovenia & 23 & 429 & 10 & 11 & 0.01 \\
4 & Smalheiser NR & University of Illinois at Chicago, USA & 21 & 5453 & 41 & 14 & 0.00 \\
5 & Swanson DR & University of Chicago, USA & 20 & 2729 & 26 & 18 & 0.00 \\
6 & Cohen T & University of Texas, USA & 19 & 1048 & 19 & 10 & 0.01 \\
7 & Song M & Yonsei University, South Korea & 18 & 889 & 15 & 6 & 0.02 \\
8 & Cestnik B & Jo\v{z}ef Stefan Institute, Slovenia & 16 & 250 & 8 & 8 & 0.00 \\
9 & Peterlin B & University Medical Center Ljubljana, Slovenia & 11 & 3438 & 28 & 7 & 0.00 \\
10 & Kastrin A & University of Ljubljana, Slovenia & 11 & 337 & 10 & 5 & 0.00 \\
\bottomrule
\end{tabularx}
\begin{tablenotes}[flushleft,para]\footnotesize
\note Rank = Ranking score based on number of publications, \textit{NP} = number of
publications, \textit{TC} = total number of citations, $h$ = $h$-index, $h_s$ =
$h$-index applied to LBD literature only, $b_i$ = betweenness centrality 
\end{tablenotes}
\end{threeparttable}
\label{tab:top_authors}
\end{table}
  
However, in terms of citations, Smalheiser scores far more than the
rest of the researchers. As expected, the most cited is his paper in which he
and Swanson as the first author described the Arrowsmith
system~\citep{Swanson1997}. Smalheiser and Peterlin also have the most
significant difference between $h$ and $h_s$ scores, meaning that they are also
highly productive outside the LBD area. The majority of authors are from the
United States (Rindflesch, Kostoff, Smalheiser, Swanson, and Cohen), four
authors come from Slovenia (Hristovski, Cestnik, Peterlin, and Kastrin), and
one from the Republic of Korea (Song). Kostoff prevails according to the
authors' dominance factor (data not shown), which is a ratio indicating the
proportion of multi-authored papers in which a person appears as the first
author. He authored 12 publications in which he appears as a first author.

The production over time was most prominent for Swanson. As stated previously,
he wrote his first paper in 1986~\citep{Swanson1986a} and the final one in
2011~\citep{Swanson2011}. Smalheiser, the researcher with the second-longest
career path, joined Swanson in 1994~\citep{Smalheiser1994}. Smalheiser
published his last paper on LBD in 2017 when he gave his personal perspective
on Swanson's contribution to science~\citep{Smalheiser2017}. In order to
understand the temporal aspects of authors' publishing patterns, we also
perform burst analysis on authors' career paths. A significant level of the
burst is presented in five authors. As expected, the founding father of LBD
had the longest burst; from 1986 until 2001. In this period Swanson published
the majority of his papers. Rindflesch's burst is significantly shorter; it
ran for six years (2011--2016). Other authors (i.e., Smalheiser, Kostoff, and
Jha) have very short bursts that last for only between one and five years.

\begin{figure}[htp]
\centering
\includegraphics[width=.7\textwidth]{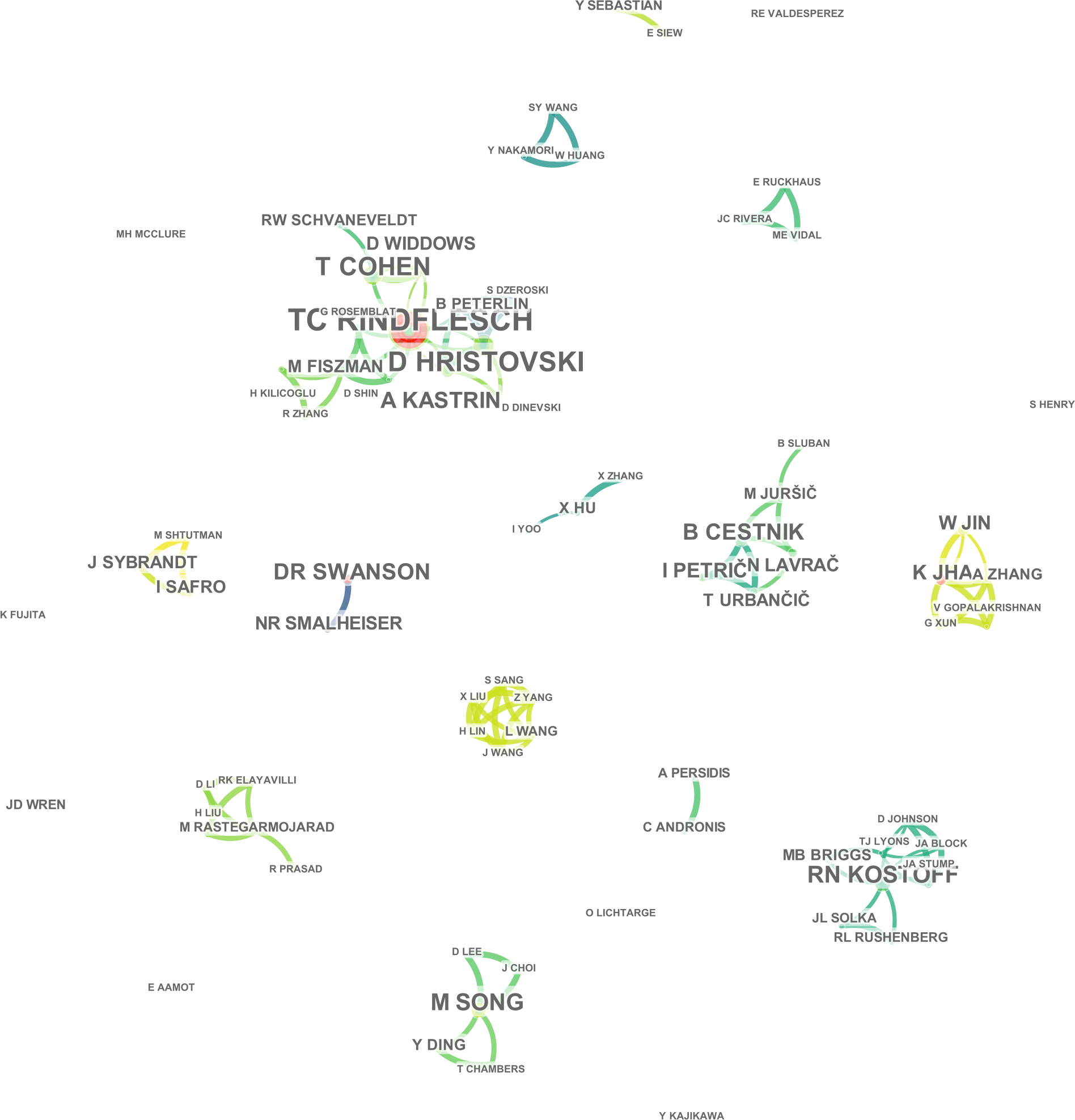} 
\caption{Co-authorship network of authors in the LBD domain. Nodes represent
  authors and the edges refer to the co-authorship relation. The size of the
  node is proportional to the number of author’s publications. The width of
  the edge is proportional to the number of co-authorships. CiteSpace
  configuration: LRF = 3, LBY = 8, e = 2.0, Top N = 50, threshold = 2.}
\label{fig:coauthorship}
\end{figure}

Let us now consider co-authorship relations. The entire co-authorship network
consists of 802 nodes and 5148 edges. Each node denotes an author and the
edges among the authors represent academic partnership through the
co-authorship on the publications. The average degree of the network was
$c = 12.84$ neighbors. The network exhibits a relatively short average path
length ($l = 4.94$ hops) and high clustering ($C = 0.72$). To draw the
network, we filter out all nodes with degree $k_i < 2$ neighbors. The reduced
co-authorship network is presented in Figure~\ref{fig:coauthorship}. The node
size is proportional to the number of publications, and the edge width follows
the strength of the collaboration. The colors of the nodes correspond to the
different network communities as identified by the Louvain clustering
algorithm~\citep{Blondel2008}. The high clustering coefficient reflects the
rich community structure of the network. The researchers inside the clusters
establish strong partnerships with colleagues in the same research group and
only weakly connect with other researchers. The biggest research community
with robust collaboration among researchers include the highly productive
research circuit of Rindflesch as the central author. The network exhibits low
density ($\rho = 0.02$), which together with high modularity ($Q = 0.91$)
indicates that research groups are dispersed. The author with the highest
betweenness centrality is Wang; however, the average betweenness centrality is
very low as well ($B = 0.001$), meaning that most authors' influence is still
at a low level. To sum up, the collaboration network exhibits many
subnetworks, indicating that the LBD domain is composed of many small and
medium-sized research groups with little communication among them.

% Authors: Top N = 50, Pruning: sliced networks, Threshold = 2, Font size = 7

\subsubsection{Countries}

\begin{figure}[htb]
\centering
\includegraphics[width=.95\columnwidth]{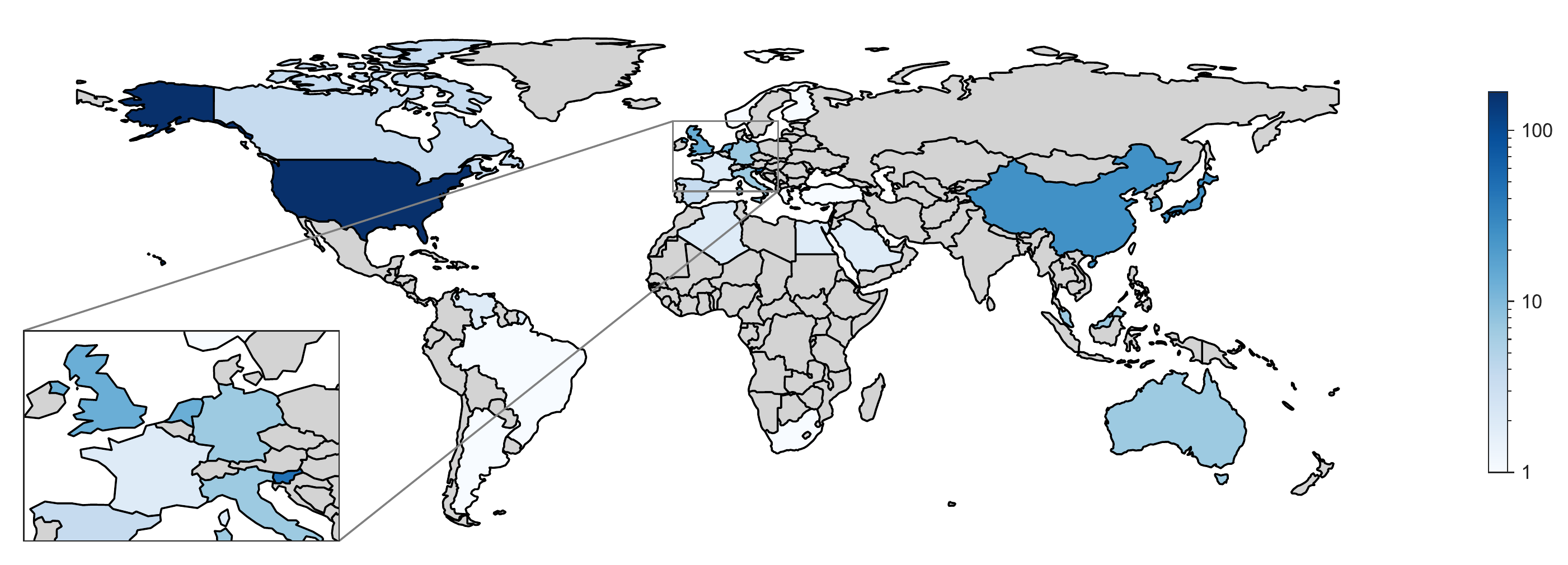}
\caption{World map of LBD research production from 1986 to 2020}
\label{fig:worldmap}
\end{figure}

A total of 27 countries have contributed to the LBD literature, as depicted in
the world map in Figure~\ref{fig:worldmap}. First, it is worth noting that LBD
production is unevenly distributed across countries. The United States commits
about half of the body of the literature to LBD research ($n = 167$ or
\SI{49}{\percent} of all the documents). This indicates that the US is leading
in LBD research. Interestingly, Slovenia, a small country in the heart of
Europe, is the second most productive country with 34 publications
(\SI{10}{\percent}). Surprisingly, India has no researcher who published about
LBD as the first author. We also report the productivity score for each
country using the simple formula (production number / population $\times$
\num{1000000}). Slovenia has been found to be the most productive country
(16.40) followed by the Netherlands (0.64) and the United States
(0.51). Following the United Nations country classification schema (\url{https://unstats.un.org}), most
countries are developed countries. Gross domestic product (GDP) measures goods
and services produced in a country. We have found a moderate positive
correlation between 2019 GDP values and number of publications of 27 countries
($r = 0.32$, $p = 0.099$). However, the United States is top-ranked according
to total citations ($n = 6267$), followed by Germany ($n = 714$), and the
Netherlands ($n = 520$).

The production concerning temporal evolution across countries reveals the
primacy of the United States. Researchers from the United States have been
publishing regularly since 1986 and they have significantly intensified the
research production since 2001. In the second place, we have identified the
United Kingdom with the publication span 1989--2019. The early appearance of
the United Kingdom on the LBD scene is due to two theoretical papers published
at the end of the 1980s that discuss the creation of new knowledge by
information retrieval~\citep{Davies1989, Davies1990}. However, the next paper
originating from the UK was published only in 2006~\citep{Song2006}. We have
detected no significant burst for countries.

\subsubsection{Institutions}

Next, we consider the institution level. Institution-based analysis might help
to discover research organizations that deserve the researcher’s attention and
provide a macro understanding of the spatial distribution of LBD efforts. Our
analysis identifies 173 different organizations that contribute to the
production of LBD publications. Please note that we only consider the
affiliation of the first author in the analysis. The details for the top 10
institutions, ranked according to the number of publications, are summarised
in Table~\ref{tab:top_institutions}. The University of Chicago stands out with
the largest number of publications ($n = 17$), thanks to the work of Swanson,
followed by the University of Illinois at Chicago ($n = 14$), and University of
Ljubljana ($n = 14$). Actually, six of the top 10 institutions come from the
United States and as many as three from Slovenia (University of Ljubljana,
Jo\v{z}ef Stefan Institute, and University of Nova Gorica). Only the University of
Chicago scores among the top 10 universities according to the Academic Ranking
of World Universities in 2019.

\begin{table}[htb]
\centering
\caption{Top 10 research institutions based on the total number of publications}
\begin{threeparttable}
\begin{tabular}{S[table-format=2.0]llS[table-format=2.0]S[table-format=4.0]c}
\toprule    
{Rank} & Institution & Country & {\textit{NP}} & {\textit{TC}} & {ARWU} \\
\midrule
1 & University of Chicago & USA & 17 & 1847 & 10 \\
2 & University of Illinois at Chicago & USA & 14 & 299 & 201--300 \\
3 & University of Ljubljana & Slovenia & 14 & 451 & 401--500 \\
4 & University of Texas & USA & 13 & 229 & 201-300 \\
5 & Office of Naval Research & USA & 12 & 480 & -- \\
6 & Drexel University & USA & 11 & 254 & 301--400 \\
7 & University of Tokyo & Japan & 11 & 78 & 25 \\
8 & Jo\v{z}ef Stefan Institute & Slovenia & 10 & 54 & -- \\
9 & National Library of Medicine & USA & 9 & 279 & -- \\
10 & University of Nova Gorica & Slovenia & 9 & 93 & -- \\
\bottomrule
\end{tabular}
\begin{tablenotes}[flushleft,para]\footnotesize
\note Rank = Ranking score based on number of publications, \textit{NP} = number of
publications, \textit{TC} = total number of citations, ARWU = Academic Ranking
of World Universities 
\end{tablenotes}
\end{threeparttable}
\label{tab:top_institutions}
\end{table}

\subsubsection{Journals}

When analyzing research productivity, it is essential to study the journals in
which papers are published. LBD is a narrow and specific research field that has
no specialized journal.\footnote{The Journal of Biomedical Discovery and
  Collaborations (DISCO) was an open access online journal that encompassed all
  aspects of scientific information management and studies of scientific
  practice. The journal connected disparate perspectives (e.g., informatics,
  computer science, sociology, cognitive psychology, scientometrics, public
  policy, technology innovation, and history and philosophy of science) and
  published several papers directly related to LBD. DISCO was published by
  BioMed Central from 2006--2008.} However, we must point out that Frontiers in
Research Metrics and Analytics recently introduced a section on Text-mining and
Literature-based Discovery which is the most noticeable development of
LBD-specific publishing venues nowadays. Instead, the LBD research is published
mainly in journals related to (biomedical) informatics and bioinformatics.
Table~\ref{tab:top_journals} summarizes details about the top 10 journals.
Interestingly, with respect to the number of publications, Lecture Notes in
Computer Science has published 21 papers on LBD research, followed by the
Journal of Biomedical Informatics with a similar number of papers.
Bioinformatics, which has the highest impact factor in our list, has published
only 7 papers on LBD. Out of 10 journals, six are published in the United
States.

\begin{table}[htb]
\centering
\caption{Top 10 journals based on the total number of publications}
\begin{threeparttable}
\begin{tabularx}{\columnwidth}{S[table-format=2.0]ZlS[table-format=2.0]S[table-format=3.0]S[table-format=1.3]}
\toprule
{Rank} & Journal title & Country & {NP} & {TC} & {IF} \\
\midrule
 1 & Lecture Notes in Computer Science & Germany & 21 & 186 & {--} \\
 2 & Journal of Biomedical Informatics & USA & 20 & 522 & 3.724 \\
 3 & Technological Forecasting and Social Change & USA & 14 & 484 & 4.040 \\
 4 & BMC Bioinformatics & UK & 13 & 423 & 2.970 \\
 5 & Information Science and Knowledge Management & USA & 11 & 70 & {--} \\
 6 & PLOS ONE & USA & 9 & 122 & 3.337 \\
 7 & AMIA Annual Symposium Proceedings & USA & 7 & 218 & {--} \\ 
 8 & Bioinformatics & UK & 7 & 346 & 8.860 \\
 9 & Journal of the American Society for Information Science and Technology & USA & 7 & 405 & 2.762 \\
 10 & Briefings in Bioinformatics & UK & 6 & 709 & 8.265 \\
\bottomrule
\end{tabularx}
\begin{tablenotes}[flushleft,para]\footnotesize
  \note NP = number of publications, TC = total number of citations, IF = 5
  year impact factor
\end{tablenotes}
\end{threeparttable}
\label{tab:top_journals}
\end{table}

\begin{figure}[htb]
\centering
\includegraphics[width=.95\columnwidth]{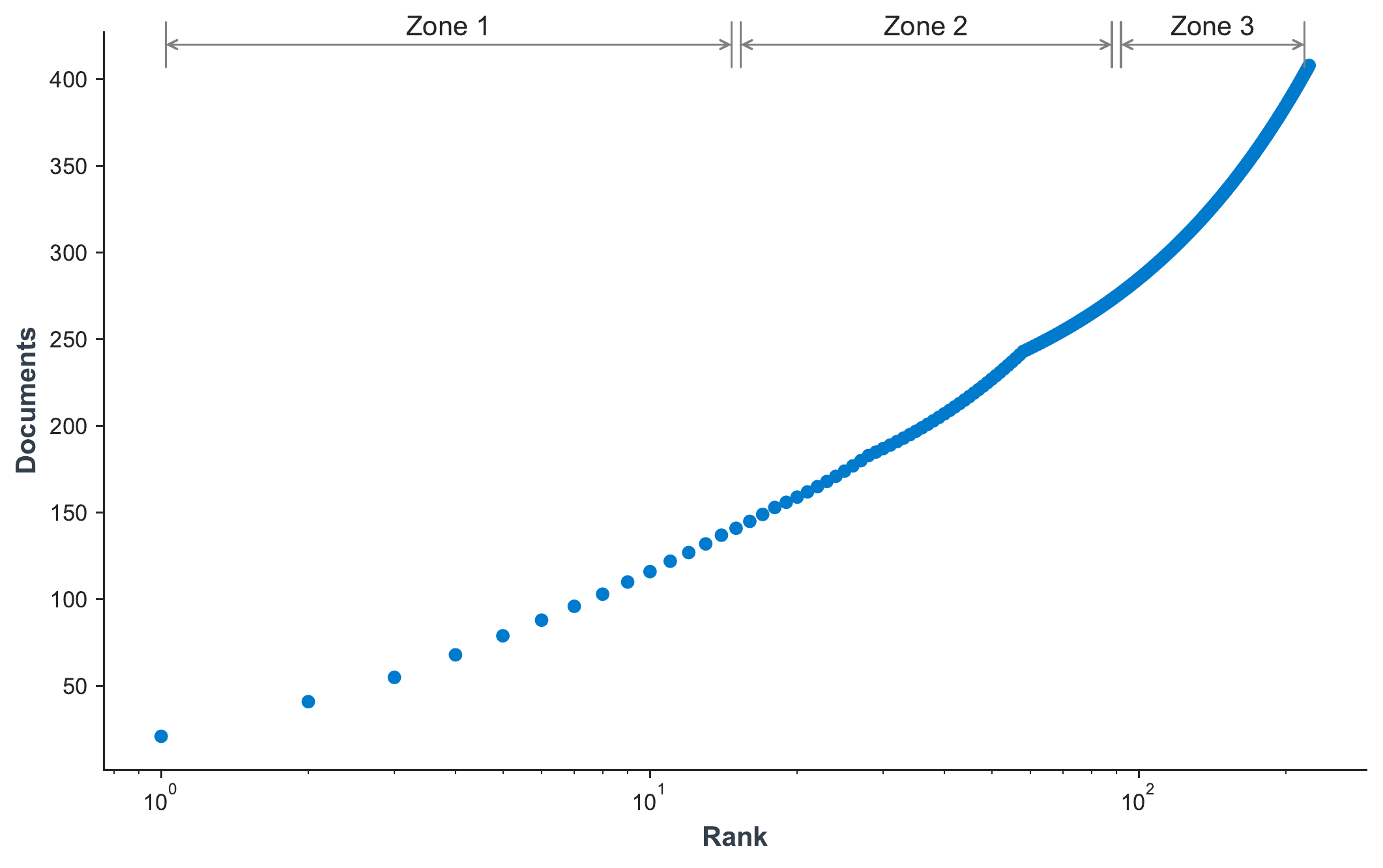}
\caption{Bradford plot. According to Bradford law, the majority of literature
  is concentrated in a small number of journals. However, our results do not
  confirm Bradford distribution. For more information please see text.}
\label{fig:bradford_distribution}
\end{figure}

To study journal distribution and to identify ``core'' journals, we have also
employed Bradford's law of scattering~\citep{Bradford1934}. In
Figure~\ref{fig:bradford_distribution} we have plotted the Bradford plot where
the cumulative frequency of LBD literature has been plotted against the
logarithm of the journal rank. The journals are grouped into three zones,
comprising a similar number of publications. The core journals are those whose
data points lie in Zone 1 ($n = 14$). The middle third (i.e., Zone 2) has 75
journals, and the last zone has 134 journals. The relationship of each zone
($14 \colon 75 \colon 134$) does not fit well into the expected Bradford's
distribution ($26 \colon 88 \colon 294$) ($\chi^2(2) = 11.17$, $p =
0.004$). Thus, we cannot confirm Bradford's assumption that a few core
journals account for one-third of all papers published within the body of LBD
literature. It is also important to note that the curve does not take a
typical ``S'' shape and there is no ``gross drop'' at the end of the curve. In
our case, the Bradford plot has taken more or less a linear shape after the
initial rise.

The main body of knowledge on LBD is ingrained in journal papers. The first
coherent book on LBD was published in 2008 by Springer and edited
by~\citet{Bruza2008}. The book contains 11 chapters by prominent authors in
the LBD field and offers the reader a comprehensive overview of LBD research.

Figure~\ref{fig:dual_map} shows a dual-map overlay visualization of LBD
publications at the journal level~\citep{Chen2014}. Each dot
represents a journal and its color indicates a cluster membership. Clusters are
labeled with the title terms of the corresponding journals. The left half of the
map refers to citing journals (i.e., the citing journal is a journal in which a
source article is published) and the right half of the map represents cited
journals (i.e., the cited journal is the journal in which a reference is
published). The spline curves connect citing journals on the left side with
cited journals on the right side.

The majority of the citing articles in Figure~\ref{fig:dual_map} are distributed
among four major disciplines, namely mathematics, biology, medicine, and health.
Furthermore, all cited papers are concentrated in molecular biology and
genetics-related journals. This finding supports the observation that LBD has
concentrated on the biomedical field.

\afterpage{
\clearpage
\thispagestyle{empty}
\newgeometry{margin=2cm}
\begin{landscape}
\begin{figure}
\centering
\includegraphics[width=\columnwidth]{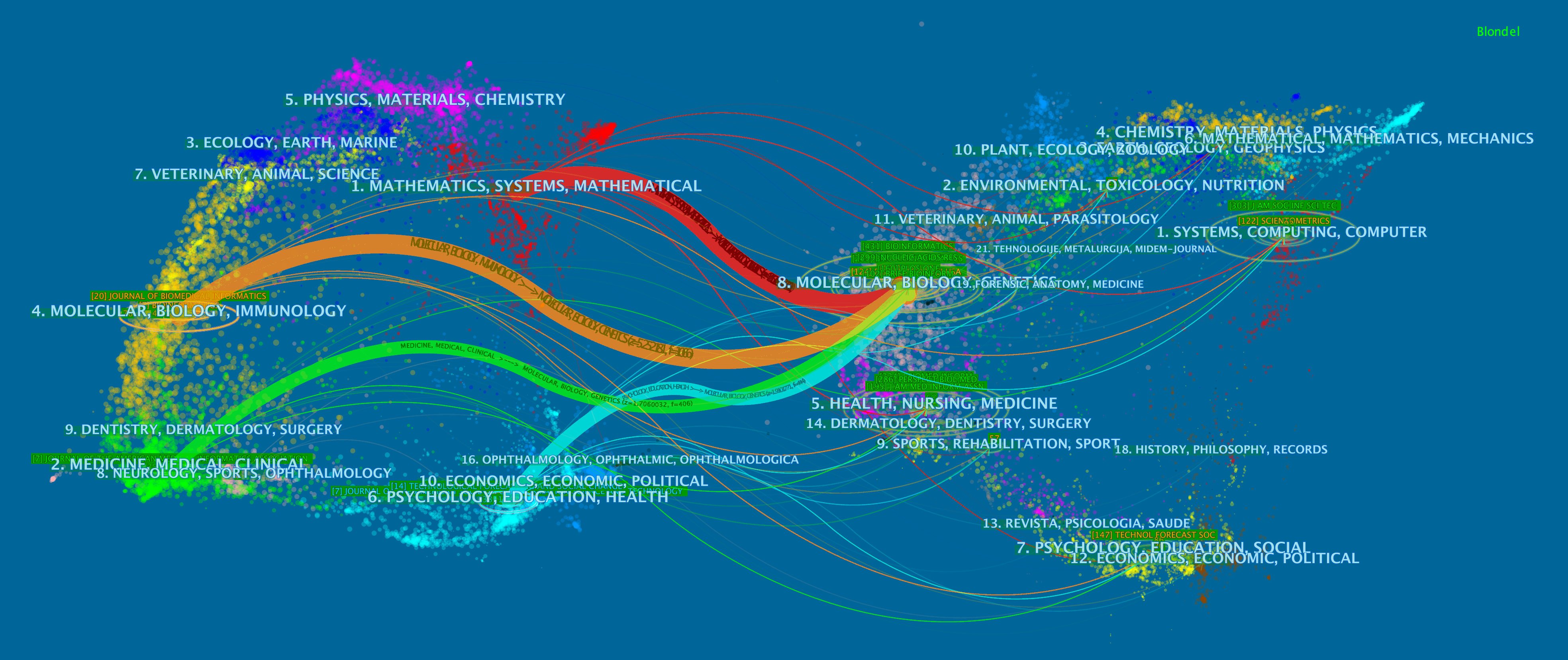} 
\caption{Dual-map overlay of journals in LBD research. The left half of the
  figure represents the citing map and the right half the cited map. The labels
  represent the disciplines covered by the journal. The lines represent the flow
  of knowledge in different scientific fields. The citation paths are bundled
  using the $z$-score function in CiteSpace.}
\label{fig:dual_map}
\end{figure}
\end{landscape}
\restoregeometry
\clearpage
}

\subsubsection{Publications}

Employing the processed bibliometric data, we can identify the most important
hallmarks of LBD research. In total, we have extracted \num{13026} references
from 409 papers related to LBD research. The top 10 most cited papers are
listed in Table~\ref{tab:top_citations}, including their first author, year of
publication, title, the total number of citations, and the number of citations
per year. The data are ranked by the number of citations. Swanson is the
author of four listed publications. First on the rank list is Swanson's
seminal paper on fish oil and Raynaud's disease~\citep{Swanson1986a}, which
has 409 citations and is cited about 12 times annually. This paper is
categorically the first hallmark of LBD research. The second and the third
rank are reserved for two review papers, written by~\citet{Cohen2005}
and~\citet{Jensen2006}. Both articles are of interest to the broader domain of
researchers because they provide an in-deep review of methods and techniques
used in text mining and especially in biomedical informatics and
bioinformatics. The most recent of highly cited papers is an interesting
article published by~\citep{Uzzi2013} in which authors discuss balancing
conventional and atypical knowledge which may be critical to link
innovativeness and scientific impact. However, it is important to note that
Uzzi's paper has considerably more citations per year in comparison to other
papers on the list. This is probably due to the high impact factor of the
Science journal in which the paper was published. A paper written
by~\citet{Chen2004} is the first serious attempt of applying a complex network
approach to LBD and is relatively highly cited among researchers who utilize
network analysis for bioinformatics. These ten publications cover the
theoretical research as well as practical applications of LBD. However, all
these papers were published before 2013, yet important scientific achievements
in LBD have also been published more recently. For example, one key
achievement that we can identify and is not on the list is a novel LBD system
called LION LBD which offers a broad range of metrics for evaluating the
strength of entity associations, and allows fast real-time discovery of
indirect associations among biomedical concepts~\citep{Pyysalo2019}. This and
other important publications are not among the current top 10 due to their
relatively recent publication date.

\afterpage{
\clearpage
%\newgeometry{margin=3cm}
%\thispagestyle{empty}
\begin{landscape}
%\vspace*{\fill}
\begin{table}
\centering
\caption{Top 10 references based on the total number of citations}
\begin{threeparttable}
\begin{tabularx}{\columnwidth}{S[table-format=2.0]lS[table-format=4.0]ZZS[table-format=3.0]S[table-format=2.2]}
\toprule
Rank & Author & {Year} & Title & Journal & {TC} & {TC/Y}     \\
\midrule

1 &%
Swanson &%
1986 &%
Fish oil, Raynaud's syndrome, and undiscovered public knowledge &%
Perspectives in Biology and Medicine &%
409 &%
11.69 \\

2 &%
Jensen et al. &%
2006 &%
Literature mining for the biologist: From information retrieval to biological
discovery &%
Nature Reviews Genetics &%
389 &%
25.93 \\

3 &%
Cohen et al. &%
2005 &%
A survey of current work in biomedical text mining &%
Briefings in Bioinformatics &%
367 &%
22.94 \\

4 &%
Kell &%
2009 &%
Iron behaving badly: Inappropriate iron chelation as a major
contributor to the aetiology of vascular and other progressive inflammatory
and degenerative diseases &
BMC Medical Genomics &%
312 &%
26.00 \\

5 &%
Uzzi et al. &%
2013 &%
Atypical combinations and scientific impact & Science &%
286 &%
35.75 \\

6 &%
Perez-Iratxeta et al. &%
2002 &%
Association of genes to genetically inherited diseases using data mining &%
Nature Genetics &%
246 &%
12.95 \\

7 &%
Swanson et al. &%
1997 &%
An interactive system for finding complementary literatures: A stimulus to
scientific discovery &%
Artificial Intelligence &%
232 &%
9.67 \\

8 &%
Swanson &%
1988 &%
Migraine and magnesium: Eleven neglected connections &%
Perspectives in Biology and Medicine &%
231 &%
7.45 \\

9 &%
Chen et al. &%
2004 &%
Content-rich biological network constructed by mining PubMed abstracts &%
BMC Bioinformatics &%
204 &%
12.00 \\

10 &%
Swanson &%
1986 &%
Undiscovered public knowledge &%
The Library Quarterly &%
171 &%
4.89 \\

\bottomrule
\end{tabularx}
\begin{tablenotes}[flushleft,para]\footnotesize
  \note TC = total number of citations, TC/Y = total number of citations per
  year
  \end{tablenotes}
\end{threeparttable}
\label{tab:top_citations}
\end{table}
%\fill
\end{landscape}
%\clearpage
%\restoregeometry
}

To gain a deeper understanding of the citation structure, we have also
conducted a burst analysis. In total, 46 documents have citation
bursts. Table~\ref{tab:top_burst} summarizes the top 10 papers with the
strongest citation burst. The top paper with the strongest burst (11.39) is
the paper written by~\citet{Swanson1997} in which they describe the Arrowsmith
discovery support system and evaluate various LBD search strategies. The
paper's burst began in 1998 and ended in 2005. The second paper with the
strongest citation burst (9.61) was written by~\citet{Weeber2001}. This paper
is one of the methodological hallmarks in the pioneering era of LBD. First, it
formally describes a two-step model of the discovery process in which research
hypotheses are generated (i.e., open discovery mode) and subsequently tested
(i.e., closed discovery mode). Second, for LBD analysis authors employ UMLS
concepts and use semantics with these concepts to filter out unmeaningful
information. Third on the list is the article by~\citet{Lindsay1999} on
lexical statistics, which initiated the era of statistically-based LBD
applications~\citep{Sebastian2017a}.

\afterpage{
\clearpage
%\newgeometry{margin=3cm}
%\thispagestyle{empty}
\begin{landscape}
%\vspace*{\fill}
\begin{table}
\centering
\caption{Top 10 references with the strongest citation bursts}
\begin{threeparttable}
\begin{tabularx}{\columnwidth}{S[table-format=2.0]lS[table-format=4.0]ZS[table-format=2.2]S[table-format=4.0]S[table-format=4.0]}
\toprule
Rank & Author & {Year} & Title & {Strength} & {Begin} & {End} \\
\midrule

1 &%
Swanson &%
1989 &%
Online search for logically-related noninteractive medical literatures: A
systematic trial-and-error strategy &%
3.67 &%
1990 &%
1997 \\
  
2 &%
Gordon et al. &%
1996 &%
Toward discovery support systems: A replication, re-examination, and extension
of Swanson's work on literature-based discovery of a connection between
Raynaud's and fish oil &%
5.27 &%
1997 &%
2004 \\

3 &%
Smalheiser et al. &%
1996 &%
Indomethacin and Alzheimer's disease &%
3.82 &%
1998 &%
2004 \\
  
4 &%
Swanson et al. &%
1997 &%
An interactive system for finding complementary literatures: A stimulus to
scientific discovery &%
11.39 &%
1998 &%
2005 \\

5 &%
Gordon et al. &%
1998 &%
Using latent semantic indexing for literature based discovery &%
4.20 &%
1999 &%
2006 \\

6 &%
Lindsay et al. &%
1999 &%
Literature-based discovery by lexical statistics &%
8.15 &%
2000 &%
2007 \\

7 &%
Smailheiser &%
1998 &%
Using ARROWSMITH: A computer-assisted approach to formulating and assessing scientific hypotheses &%
4.94 &%
2001 &%
2006 \\

8 &%
Weeber et al. &%
2001 &%
Using concepts in literature-based discovery: Simulating Swanson's
Raynaud-fish oil and migraine-magnesium discoveries &%
9.61 &%
2003 &%
2006 \\

9 &%
Weeber et al. &%
2000 &%
Text-based discovery in biomedicine: The architecture of the DAD-system &%
5.56 &%
2003 &%
2005 \\                                                                  

10 &%
Swanson et al. &%
1999 &%
Implicit text linkages between Medline records: Using Arrowsmith as an aid to
scientific discovery &%
4.16 &%
2003 &%
2006 \\
  
\bottomrule
\end{tabularx}
\begin{tablenotes}[flushleft,para]\footnotesize
\note Strength = strength of burst, Begin = begin of burst, End = endo of burst
  \end{tablenotes}
\end{threeparttable}
\label{tab:top_burst}
\end{table}
%\fill
\end{landscape}
\clearpage
%\restoregeometry
}

\subsection{Science mapping}

Science mapping is the next logical step in our analysis. In this section we
provide keywords analysis and DCA.

\subsubsection{Keyword analysis}

\begin{figure}
\centering
\includegraphics[width=.85\columnwidth]{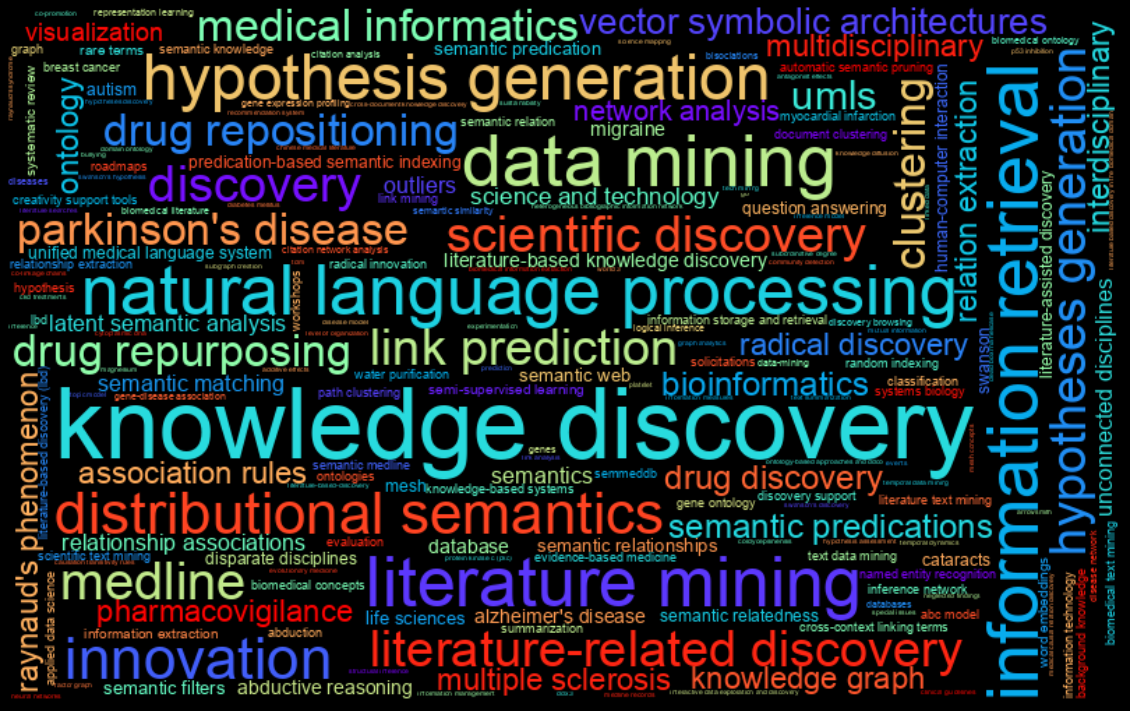}
\caption{Word cloud of keywords extracted from documents on LBD
  research. Please note that the term \textit{literature-based discovery} is
  omitted.}
\label{fig:wordcloud}
\end{figure}

The research topics studied in LBD research can be characterized by the
keywords assigned to each bibliographic record. The keywords enable us to
summarise, qualify, and explain the entire scientific document within the
boundaries of a particular research domain. The keywords provide a plausible
summarization of research hotspots, for example, burst keywords represent
research frontiers and indicate possible emerging trends. The word cloud in
Figure~\ref{fig:wordcloud} reflects the frequency distribution of keywords in
the core set of 409 documents. The five most frequent keywords are
\textit{knowledge discovery}, \textit{information retrieval}, \textit{data
  mining}, \textit{natural language processing}, and \textit{literature
  mining}. It is important to note that we detect no significant burst among
the keywords.

%\afterpage{
%\clearpage
%\clearpage
%}

The timeline view of the keywords is presented in
Figure~\ref{fig:timeline}. The timeline starts with the year 1996 because
keywords were rarely assigned to bibliographic records before this early
period. The research until the year 2001 is topical and domain-specific; the
timeline is loaded with keywords such as \textit{fish oil},
\textit{information retrieval}, \textit{raynaud}, \textit{magnesium}, and
\textit{migraine}. Entering the new millennium, the richness of keywords
increases, and terms evolve rapidly. In the 2010s we can observe two main
directions of research: one is interweaving of LBD ideas with genetics; the
other concerns application of semantics in LBD. Prevailing keywords are
\textit{text mining}, \textit{knowledge discovery}, \textit{data mining},
\textit{disease}, \textit{information}, \textit{system}, \textit{gene}, and
\textit{natural language processing}. This indicates that LBD research in this
decade was developing from baseline Swanson's approach and its replications
(e.g.,~\citet{Gordon1996}) towards systematic development of knowledge
discovery methods and data mining tools for LBD. In terms of LBD reviews from
this era we can observe a shift from statistical-based approaches (e.g., pure
co-occurrence) to rule-based (e.g., association rules) and semantic
approaches. For example, in 2005~\citet{Hristovski2005} introduced the BITOLA
system, which utilizes association rule mining to reveal (gene-disease)
relations between biomedical concepts by observing frequent patterns among
data objects. Roughly at the same time, researchers introduced a
semantic-based discovery pattern approach~\citep{Hristovski2006, Ahlers2007},
which significantly increases the precision and enhances the interpretability
of LBD systems. A plethora of other Web tools and services were also developed
within this decade (e.g., DAD~\citep{Weeber2001}, LitLinker~\citep{Pratt2003},
Manjal~\citep{Srinivasan2004}, IRIDESCENT~\citep{Wren2004b}, and
RaJoLink~\citep{Petric2009}). The fourth decade of LBD (i.e., 2011--2020) has
been the decade of network science in the LBD community. Important keywords in
this time period are \textit{network}, \textit{link discovery}, \textit{link
  prediction}, \textit{network analysis}, \textit{knowledge graph},
\textit{drug discovery}, and \textit{pharmacovigilance}. According
to~\citet{Sebastian2017a} this decade coincides with the stage of emerging LBD
approaches which could be characterized by two directions. First, traditional
co-occurrence-based and knowledge-driven approaches culminate in solutions
that integrate both. Second, LBD becomes more interdisciplinary, incorporating
methods and tools from information sciences~\citep{Chen2009b},
scientometrics~\citep{Kostoff2014}, and machine
learning~\citep{Sebastian2017b}.

\afterpage{
\clearpage
\thispagestyle{empty}
\newgeometry{margin=1cm}
\begin{landscape}
\begin{figure}
\centering
\includegraphics[scale=.12]{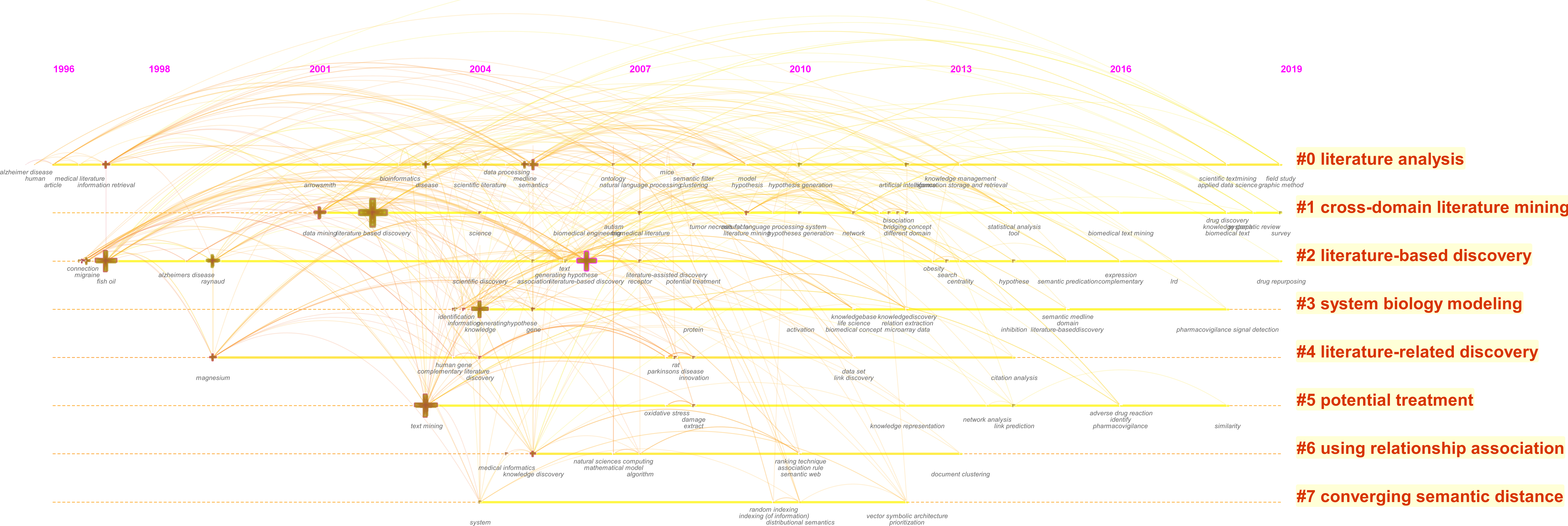} 
\caption{A timeline view of keywords extracted from documents on LBD for the
  period 1986--2020. We have built a list of keywords from author keywords and
  keywords assigned by database curators. For the description of the clusters
  please see the text. Online version of the figure
  (\url{https://doi.org/10.5281/zenodo.3884422}) can be enlarged for further
  details.}
\label{fig:timeline}
\end{figure}
\end{landscape}
\restoregeometry
\clearpage
}

\subsubsection{Document co-citation analysis}

DCA allows us to examine a network of co-cited references. The body of cited
references provides the knowledge base of the selected
documents. Figure~\ref{fig:dca} depicts the subset of the DCA network as
pruned with the Pathfinder algorithm~\citep{Chen2006}. The network exhibits
399 nodes and 891 edges. Each node in the network refers to a document that is
labeled with the first author name and the year of publication. Each edge
represents a co-citation relation among the pair of documents. The size of the
nodes is proportional to the co-citation frequency. The most highly cited
documents were written by~\citet{Srinivasan2004} and \citet{Wren2004a}. As we
have said previously in the Methods section, we have computed two types of
node importance, namely betweenness centrality, and brokerage. The
aforementioned Wren's paper is the document with the highest betweenness
centrality (0.34), followed by~\citet{Jensen2006} (0.27). Together
with~\citet{Frijters2010}, both papers also exhibit the highest brokerage
score. These references are not only important as hubs in the DCA network but
also as bridging nodes among contextually different groups of nodes.

\afterpage{
%\clearpage
\begin{figure}[p]
\centering
\includegraphics[width=.8\textwidth]{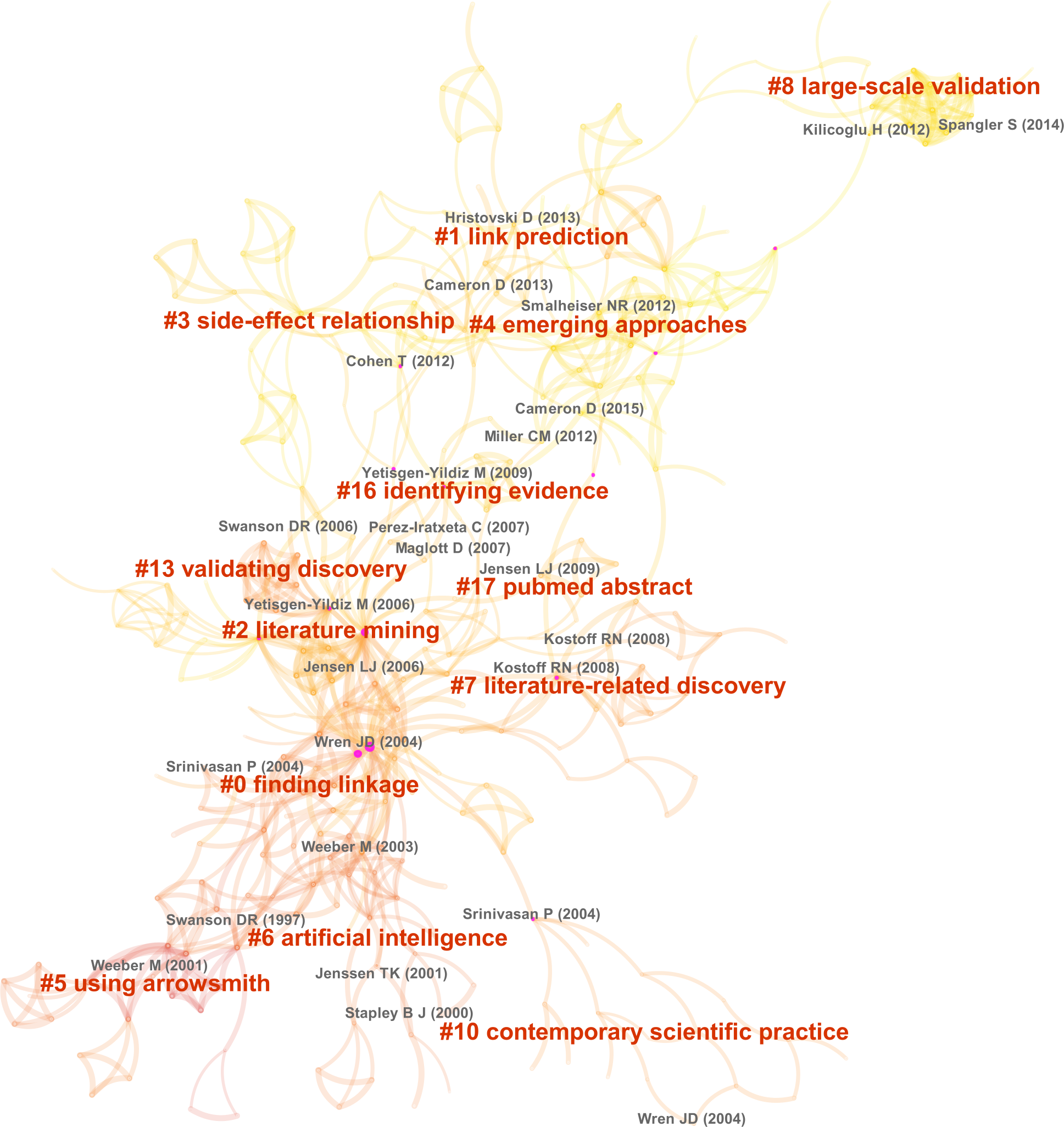} 
\caption{Document co-citation network of the core dataset based on documents
  published between 1986 and 2020. Cluster labels in red text are taken from
  the titles of the cited documents using the log-likelihood ratio
  algorithm. Red nodes refer to documents with high citation burst. For the
  description of the clusters please see the text. CiteSpace configuration:
  LRF = 3, LBY = 8, e = 2.0, Top N = 50.}
\label{fig:dca}
\end{figure}
%\clearpage
}

Using cluster analysis of the cited references we obtain a set of 44
co-citation clusters that may provide us the main research topics of the
intellectual base. Figure~\ref{fig:dca} depicts the DCA network with the
embedded clusters. In total we have identified 13 clusters that are worth
further consideration. Table~\ref{tab:dca_clusters} summarizes the basic
statistics for each cluster sorted by its size: ID number, size, silhouette
width, mean year, and first cluster label as identified by the log-likelihood
ratio extraction method. The silhouette width ranges from 0.72 to 0.99
indicating adequate consistency of derived clusters. The clusters could be
summarized as follows.
\begin{enumerate}
\item Cluster \#0 (\textit{finding linkage}) has the largest number of
  members. Cluster's mean age is 2003 and is relatively old. This cluster is
  to be interpreted as a general LBD cluster while it contains generic phrases
  such as \textit{life science}, \textit{online tool}, \textit{biomedical text
    mining}, \textit{vector space model}, and \textit{discovery
    approach}. Representative references in this cluster
  are~\citet{Lindsay1999}, \citet{Weeber2003}, and \citet{Hristovski2005}
  which also exhibit the highest citation burst. The paper written by
  Hristovski et al. demonstrated high Sigma\footnote{Sigma ($\Sigma$) index is
    used to characterize scientific novelty according to centrality and
    burstness as criteria of transformative discovery~\cite{Chen2009b}. Sigma
    is defined as $(\textit{centrality} + 1)^{\textit{burstness}}$.} value,
  indicating a high degree of scientific novelty. Two typical citing papers
  are for example Cohen's~\citeyearpar{Cohen2005} review on text mining with a
  special section on LBD and Weeber et al.'s~\citeyearpar{Weeber2005} paper,
  in which the authors review Web-based tools for LBD.
\item Cluster \#1 (\textit{link prediction}) is much younger; its mean age is
  2011 and it assembles 42 references. The main theme of this cluster is on
  the prediction of future discoveries using link prediction methods. The
  cluster is loaded with terms reflecting its affinity to complex networks
  science (e.g., \textit{semantic medline network}, \textit{mesh co-occurrence
    network}, \textit{supervised link discovery}). The three top-cited
  references are~\citet{Cameron2015}, \citet{Hristovski2013},
  and~\citet{Wilkowski2011}. Representative citing papers include seminal work
  by~\citet{Katukuri2012} on supervised link discovery and Kastrin et
  al.'s~\citeyearpar{Kastrin2016} generalization of link prediction for LBD.
\item Cluster \#2 (\textit{literature mining}) contains 35 members and refers
  to gene prioritization and drug repurposing using LBD methods. It contains
  phrases such as \textit{high-throughput literature analysis},
  \textit{disease candidate gene}, and \textit{gene prioritization}. The three
  most representative references for this cluster are \citet{Wren2004a},
  \citet{Frijters2010}, and \citet{Jensen2006}. The latter is a highly cited
  review paper on biomedical text mining and exhibits high Sigma value,
  reflecting its novelty in the field. Two typical citing articles are for
  instance Andronis's~\citeyearpar{Andronis2011} paper on literature mining
  for drug repositioning and Deftereos's~\citeyearpar{Deftereos2011} review on
  adverse event prediction using literature analysis.
\item Cluster \#3 (\textit{side-effect relationship}) consists of 32
  members. The representative phrases in this cluster are both technical
  (e.g., \textit{learning predictive models}, \textit{literature-derived
    semantic predication}) as well as applied (e.g., \textit{large clinical
    dataset}, \textit{identifying plausible adverse drug reactions},
  \textit{adverse event prediction}). Some representative cited references
  with the highest citation bursts are \citet{Cohen2012}, \citet{Shang2014},
  and \citet{Cameron2013}. The citing papers include for example Cohen’s
  papers in which he discusses and elaborates the methodology of embedding of
  semantic predications~\citep{Widdows2015, Cohen2017}.
\item Cluster \#4 (\textit{emerging approaches}) consists of 28 members. The
  average mean year of 2013 reflects its relative recentness. This cluster is
  mainly related to the description of novel and emerging approaches in LBD,
  which is reflected in phrases and terms such as \textit{new approach},
  \textit{heterogeneous bibliographic information network}, or
  \textit{convolutional neural network method}. The most cited reference with
  the highest citation burst is~\citet{Smalheiser2012}. The citing articles
  include recent reviews by \citet{Sebastian2017a}, \citet{Smalheiser2017},
  and \citet{Thilakaratne2019b}.
\item Cluster \#5 (\textit{using arrowsmith}) is the oldest extracted cluster
  (with a mean age of 1998). It reflects early pioneering days of LBD. Some
  representative phrases are for instance \textit{lexical statistics},
  \textit{human-computer collaboration}, and \textit{medline record}. In this
  cluster, the most representative cited references with the highest burst are
  \citet{Swanson1997}, \citet{Weeber2001}, and~\citet{Gordon1998}. The
  representative citing articles include Swanson and Smalheiser's
  \citeyearpar{Swanson1999} paper on using Arrowsmith for biomedical relation
  discovery and Lindsay and Gordon's \citeyearpar{Lindsay1999} article in
  which they applied lexical statistics to extend and replicate original
  Swanson's discoveries.
\item Cluster \#6 (\textit{artificial intelligence}) contains 22 members and
  is represented with terms such as \textit{emerging in-silico scientist},
  \textit{bridging biology}, and \textit{content-rich biological network}. The
  most cited and highly burst references in this cluster are
  \citet{Jenssen2001}, \citet{Stapley2000}, and~\citet{Hristovski2001}. Citing
  papers include for instance Chen and Sharp's \citeyearpar{Chen2004} paper on
  building biological networks beyond pure co-occurrence approach.
\item Cluster \#7 (\textit{literature-related discovery}) may be referred to
  as Kostoff's cluster. Its mean year is 2006 and is loaded with terms like
  \textit{potential treatment} or \textit{future research directions}. This
  cluster contains mainly Kostoff's references that were part of the special
  issue of the journal Technological Forecasting and Social Change in 2008.
\item Cluster \#8 (\textit{large-scale validation}) is the youngest cluster in
  this list (mean year = 2014). It contains 20 members and is loaded with
  phrases such as \textit{hypothesis generation system}, \textit{candidate
    ranking}, and \textit{automated literature mining}. One and only reference
  with a significant burst in this cluster is Kilicoglu et
  al.'s~\citeyearpar{Kilicoglu2012} paper on SemMedDB, a large scale
  repository of semantic predication extracted from MEDLINE. The citing papers
  reflect the recent work on large-scale LBD by~\citet{Sybrandt2018a,
    Sybrandt2018b}.
\item Clusters \#9, \#13, \#16, and \#17 are the smallest ones. Each of them
  consists of less than 20 members and are thus unstable to interpret.
\end{enumerate}

\begin{table}[htb]
\centering
\caption{Co-citation clusters of LBD research 1986--2020}
\begin{threeparttable}
\begin{tabular}{S[table-format=2.0]S[table-format=2.0]S[table-format=1.2]S[table-format=4.0]l}
\toprule   
{ID} & {Size} & {Width} & {Year} & Label \\
\midrule
0  & 43 & 0.72 & 2003 & finding linkage \\
1  & 42 & 0.83 & 2011 & link prediction \\
2  & 35 & 0.82 & 2008 & literature mining \\
3  & 32 & 0.87 & 2011 & side-effect relationship \\
4  & 28 & 0.86 & 2013 & emerging approaches \\
5  & 24 & 0.96 & 1998 & using arrowsmith \\
6  & 22 & 0.92 & 2001 & artificial intelligence \\
7  & 21 & 0.97 & 2006 & literature-related discovery \\
8  & 20 & 0.99 & 2014 & large-scale validation \\
9  & 17 & 0.99 & 2003 & contemporary scientific practice \\
13 &  9 & 0.97 & 2002 & validating discovery \\
16 &  7 & 0.96 & 2005 & pubmed abstract \\
17 &  7 & 0.95 & 2011 & identifying evidence \\
\bottomrule
\end{tabular}
\begin{tablenotes}[flushleft,para]\footnotesize
  \note ID = cluster ID, Size = number of references in a cluster, Width =
  silhouette width of a cluster, Year = mean year, Label = cluster label as
  identified by a LLR algorithm
\end{tablenotes}
\end{threeparttable}
\label{tab:dca_clusters}
\end{table}

\subsubsection{Cascading citation expansion}

As described in the Methods section, the basic idea of the cascade citation
expansion is to select a seed article and iteratively expand the initial set of
references by adding new papers through citation links~\citep{Chen2019}. We used
Swanson's paper on fish oil and Raynaud's disease~\citep{Swanson1986a} as a seed
article, which is generally considered to be a pioneering and groundbreaking
work in LBD~\citep{Chen2019, Gopalakrishnan2019, Henry2017, Sebastian2017a,
  Smalheiser2012, Smalheiser2017, Thilakaratne2019a, Thilakaratne2019b}. For
testing purposes, we used 2- and 3-generation forward expansion. After manual
inspection, we decided to present the results only for a 2-generation procedure,
which optimally summarizes the tradeoff between the recall of relevant papers
and their precision.

The 2-generation forward expansion procedure collects a total of \num{86927}
distinct references. After pruning with the Pathfinder algorithm, we obtain the
merged network with 622 nodes and 1267 edges. The extracted network contains 478
nodes, which is \SI{76}{\percent} of the full network we have obtained using the
expansion procedure. Figure~\ref{fig:dca_expansion} depicts the largest
connected component of the co-citation network.

\afterpage{
\clearpage
\begin{figure}[p]
\centering
\includegraphics[width=.8\textwidth]{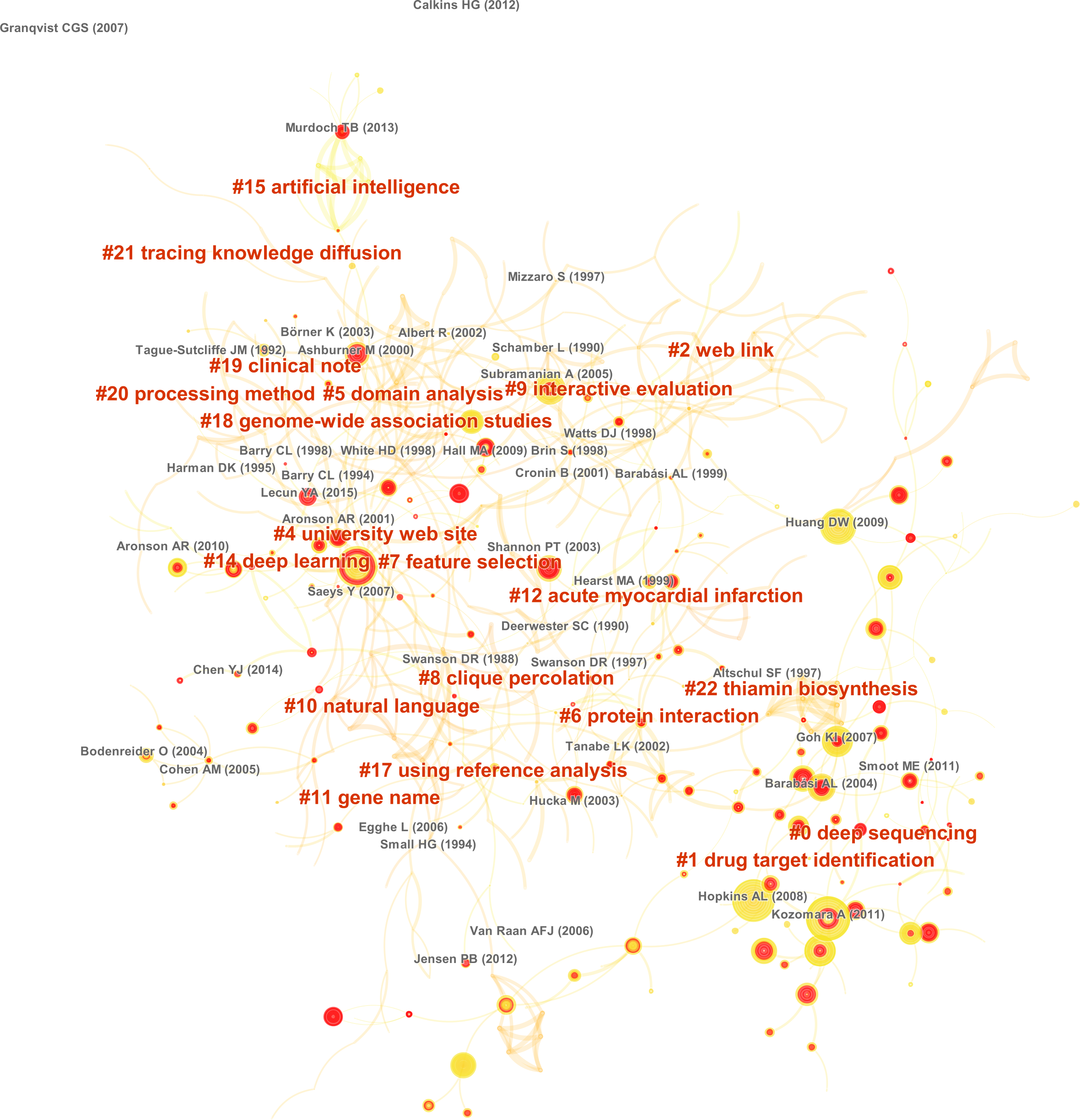} 
\caption{Document co-citation network derived from a 2-generation forward
  citation expansion procedure from a seed article by Swanson
  procedure. Clusters are labeled in red text. Red nodes indicate documents
  with high citation bursts. For the description of the clusters please see
  the text. CiteSpace configuration: LRF = 3, LBY = 10, e = 2.0, Top N = 50.}
\label{fig:dca_expansion}
\end{figure}
\clearpage
}

Most cited is the paper by \citet{Kozomara2011} which introduced the miRBase,
a primary Web repository for microRNA sequences and annotations. Our
inspection reveals that RNA research and LBD are related through Smalheiser’s
clinical work. For a deeper understanding of this connection, we refer the
reader to the paper by \citet{Chen2019}, who deduce similar conclusions. A
citation burst has been detected in 227 papers. Among the top 25 papers with
the strongest citation burst we identify three of Swanson's works including
his seminal paper on fish oil and Raynaud's disease~\citep{Swanson1986a} and a
subsequent paper on migraine and magnesium~\citep{Swanson1988}.

The modularity of the network is high ($Q = 0.91$). Clustering reveals 68
coherent groups of nodes, out of which 20 significant clusters are labeled in
Figure~\ref{fig:dca_expansion}. All silhouette widths are in the range
0.87--0.99 indicating high homogeneity of clusters. The oldest cluster is \#10
(natural language) with 1990 as a mean year of publication. Most interesting
are the youngest clusters that might indicate the new trends and topics which
are worth addressing in LBD research. The youngest (mean age = 2014) are
clusters \#14 (deep learning) and \#15 (artificial intelligence). The former
includes terms like convolutional neural network, reinforcement learning, and
machine learning, while the latter contains keywords such as big data
analytics, computational intelligence, and precision medicine.

Until recently, neural network models have been rarely used in LBD
applications~\citep{Crichton2018, Sang2018, Crichton2020}. Although they have
great potential to achieve better prediction performance and more stable
results than ABC-based methods, their output suffers from low explainability
and interpretability, due to their black-box nature~\citep{Zitnik2019}. But on
the other hand, this is also a great opportunity for artificial
intelligence. Developing prediction models with the ability to explain the
statistical learning process is currently a hot topic trend under the umbrella
of the so-called Explainable Artificial Intelligence
(XAI)~\citep{BarredoArrieta2020}. Thus, the combination of both deep learning
models and XAI presents fundamentally new frontiers for LBD research.

\section{Discussion}

LBD research is more than thirty years old. In this work, we have conducted a
scientometric analysis that provides a detailed overview of the LBD
literature, its intellectual structure, and dynamics. The present work
demonstrates that the publication trend is increasing in the LBD
community. Our investigations show a colorful palette of authors and topics in
the various nuances of LBD research. The findings offer insights on the
current state of the LBD research and provide future research directions such
as deep learning and XAI. The current study also extends the current classical
reviews on LBD. To the best of our knowledge, this is the first inclusive
scientometric analysis of the body of research evidence in LBD.

Understanding the past and the current body of publications is a sine qua non
for growing the LBD research in the future. In the recent decade, there have
been a plethora of studies examining knowledge structure and evolution through
the scientometric lens of particular scientific fields. The lack of similar
studies in the LBD area makes it difficult if not impossible to compare LBD
with other research fields. However, LBD leans to biomedicine and medical
informatics in particular. There are two reasons for this fact. First,
historically, LBD originates from medical applications~\citep{Swanson1986a,
  Swanson1988}. Second, in a practical sense, the MEDLINE distribution is
freely available to researchers, which is not the case with WoS or Scopus. Due
to its availability, a number of open-source Web applications that use MEDLINE
and text mining for the purpose of LBD have been
developed~\citep{Gopalakrishnan2019}. A review of empirical evidence reveals
that several bibliometric studies have been performed in the domain of
biomedical informatics. For example, \citet{DeShazo2009} characterized the
field of medical informatics in general over a 20 year period
(1987--2006). \citet{Schuemie2009} identified three main subdomains of medical
informatics including health information systems, knowledge representation,
and data analysis. Last but not least, \citet{Nadri2017} performed
bibliometrics analysis on the top 100 most cited papers in medical informatics
and demonstrated the dominance of statistics and artificial intelligence
sub-areas.

Relative to other research fields, LBD can be considered young. However, from
its beginnings in the mid-1980's it has grown into a mature scientific
discipline, even with its own entry in the MeSH vocabulary. We have
demonstrated that LBD interacts with various research fields, so it is
needless to say that LBD stands on the shoulders of many giants, including
information science, text mining, and natural language processing. Even more,
our hypothesis is that LBD effectively adopts and recycles the ideas and
research trends from other research fields, such as natural language
processing and statistics. The first such example was the research conducted
by~\citet{Gordon1998}, who borrowed latent semantic
analysis~\citep{Deerwester1990} to compute the semantic similarity between a
source term and a target term in the LBD discovery process. Recently,
\citet{Lever2017} showed that singular-value decomposition, a well-known
factorization method from linear algebra, provides the best scoring approach
for predicting future co-occurrences in comparison to the leading methods in
LBD. This behavior is in line with the theory of transformative discoveries
proposed by \citet{Chen2009b}. The central postulate of this theory is that
connecting otherwise divergent pieces of knowledge is an important mechanism
of creative thinking in science. In addition, \citet{Uzzi2013} empirically
demonstrated that the highest-impact science is based on unusual combinations
of existing patches of knowledge.

A conspicuous change in the number of papers published per year suggests that
a major turning point is occurring in the field. We have found that the number
of publications has been increasing over the last 20 years, particularly since
2008. The development of the LBD field is associated with great progress in
computer science and natural language processing in particular. It is
important to note that there are some general factors promoting the
development of the field, such as knowledge fragmentation, overspecialization,
and information overload. On the other hand, availability of text mining
resources (e.g., SemMedDB~\citep{Kilicoglu2012}, PubTator
Central~\citep{Wei2019}) has allowed more people to work on the LBD
problem. The total citations accumulate over the years and consequently, the
recent papers do not have enough time to acquire more citations. The growth of
publications and citations in the last decade indicates a promising future of
LBD. According to Shneider's~\citeyearpar{Shneider2009} four-stage theory of
scientific evolution, the research process is classified into four phases:
\begin{enumerate*}[label=(\roman*)]
\item the first phase introduces new subject matters into the realm of
  science;
\item the second phase develops domain methodology, enabling the language to
  describe a broad spectrum of phenomena;
\item the third phase applies known research methods to new research subject
  matters; and
\item the last, fourth, phase records existing knowledge and puts it into
  practical use.
\end{enumerate*}
In line with the above categorization, LBD could be placed in the second,
tool-construction, stage. In the first stage, pioneers such as Swanson and
Smalheiser conceived the field, identified central research questions, and
built first applications. In the second stage, the LBD community built and
improved computation tools to systematically study original problems. However,
we are still far away from the fourth stage. In our opinion, we need to bring
together efforts from both the computer science community and biomedical
literature mining groups. While the former is oriented towards developing new
algorithms, the latter tends more to solve applied problems~\citep{Zhao2020}.

Scientific productivity is strongly correlated with international
collaboration among researchers, countries, and
institutions~\citep{Lee2005}. Studies investigating the scientific impact of
cross-institution groups confirm that their papers have a higher citation rate
in comparison to papers produced by a single research group. Papers with
international co-authorship have an even higher
impact~\citep{Thonon2015}. However, we have demonstrated that there is a lack
of collaboration among research groups in the field of LBD. Most of the
research produced in the field of LBD is generated by small cliques of
researchers. Even though the collaboration and internationalization among
researchers have certain, mainly societal, downsides, it provides great
benefits. \citet{Abramo2011} demonstrated an increasing trend in collaboration
among institutions could be attributed to different policies stimulating
research collaboration (e.g., the EU Framework Programme for Research and
Innovation). We are aware of at least one successful EU FP7 funded project
from the broad domain of LBD named BISON (2008--2011) that investigates novel
methods for discovering new, domain-bridging connections and patterns from
heterogeneous data sources~\citep{Berthold2012}.

Although the number of publications has grown over the years, scientific
production in LBD is still very limited and evolves much slower than
comparable research fields (e.g., scientometrics~\citep{Hou2018}). It is worth
noting that the scientific production in the LBD field across countries is far
from uniform. Among the 10 countries with the greatest contribution to the
field, the United States have contributed most papers, far exceeding other
countries. This is not surprising as they have more established research
backgrounds and research funding.

It is known that today's science stimulates the growth of large teams in all
areas of research, whereas small teams and solitary researchers
diminish~\citep{Wu2019}. Small groups disrupt science with new ideas,
concepts, and theories, while large teams tend to further develop existing
ones. Our results have demonstrated that LBD research is distinctively
partitioned into small groups. This is somehow surprising, since modern LBD
problems are highly complex and require interdisciplinary work (i.e., a team
of information scientists, computer scientists, statisticians, natural
language processing experts, etc.). On the other hand, \citet{Li2019b} show
that exceptional scientific achievement comes from small teams. However, we
agree with~\citet{Wu2019} that to further strengthen the science and LBD, in
particular, it is necessary to amplify both
\begin{enumerate*}[label=(\roman*)]
\item science disruption by exploring older and lesser known but promising
  work (such research was for example demonstrated by~\citet{Lever2017}, who
  combined LBD and singular value decomposition) and
\item solving already known problems and refining common designs (such
  research was for instance demonstrated recently by~\citet{Crichton2018}, who
  used neural link prediction for LBD).
\end{enumerate*}
A possible solution might be to organize an international scientific
conference\footnote{To our knowledge two LBD events have been organized in the
  past decade. The First International Workshop on the role of Semantic Web in
  Literature-Based Discovery (SWLBD 2012) was co-organized with The IEEE
  International Conference on Bioinformatics and Biomedicine in Philadelphia,
  USA (\url{http://www.ischool.drexel.edu/ieeebibm/bibm12}). At the time of this
  writing, Smalheiser and Sebastian organized the First International Workshop
  on Literature-Based Discovery (LBD2020) co-located with the 24-th
  Pacific-Asia Conference on Knowledge Discovery and Data Mining in Singapore
  (\url{http://scientificarbitrage.com/lbd-2020}).} (or some other sort of
capacity building) dedicated solely to LBD, which is needed to facilitate
networking among LBD researchers. Such an event might significantly contribute
to the further development of top-level research and foster collaboration,
especially between researchers from different countries.

A keyword analysis is a frequently used technique in scientometrics to outline
preferences, trends, and emerging tendencies in a set of publications. Our
exploration reveals the tight integration of LBD research with information
retrieval and natural language processing themes. On a higher semantic level,
LBD relates to computer science and computational biology in
particular. Through the whole keyword network, we can observe the intertwining
of technical (e.g., artificial intelligence, text mining) and biomedical
(e.g., gene, genetics) terms. This network is constructed using the authors’
keywords and terms attributed by the database curators. For presentation
purposes, we need to greatly reduce the network. This way we lose
low-frequency terms which may indicate new, emerging trends in LBD. However,
manual exploration of the LBD literature has identified deep learning as a
possible future direction in LBD. Two main contributions were published about
LBD and deep learning at the end of 2018: an article written by Korhonen's
lab\footnote{At the time of the submission of this paper we came across a new
  paper~\citep{Crichton2020} from Korhonen's group which discussed
  implementation of graph-based neural network methodology for open and closed
  LBD. }~\citep{Crichton2018} and a short, only two-page long conference paper
by~\citet{Sang2018}.

Despite its contributions, this study also has certain limitations. First,
although our review is based on deliberate search queries in eight most
comprehensive databases, it may be that other search strategies have yielded
(slightly) different results. However, it is very difficult to define a search
query that covers all the relevant papers in the scientific literature while
simultaneously excluding irrelevant papers. Second, employed databases
preferably list English-language publications. To this end, some papers in
other languages might not have been included. Third, the presented approach
uses solely quantitative methods, without in-depth qualitative interpretation
of the content. Despite these limitations, we believe that we properly present
a worldwide view on LBD in the last four decades. Additional future work
should also consider combining quantitative and qualitative analysis to
further extend this analysis.

\section{Conclusions}

In this study, we have performed a comprehensive study of the worldwide
scientific output of LBD research from 1986 to 2020. During this time span,
the number of LBD publications increased. Rindflesch published the highest
number of publications on LBD, followed by Swanson. Swanson was also the most
cited author in the field. The United States was the most dominant country
according to the number of published papers. The University of Chicago was the
most influential institution, both with the largest number of publications on
LBD, as well as with the largest number of citations.

In the future, we anticipate that studies in LBD extend in two directions,
namely deep learning and XAI. Both fields are strongly connected to advanced
machine learning techniques and next-generation network science including
network embeddings. For example, graph neural networks provide a very powerful
toolbox, achieving excellent performance on a wide scope of tasks including
node classification and link prediction~\citep{Ying2019}. On the other hand,
we need to ensure strong interpretability and explainability of prediction
models, while black-box nature of the current deep learning models often
limits their adoption in practical applications.

To sum up, this study could significantly augment the traditional literature
reviews and provide helpful information to determine new research directions
and perspectives of LBD research. More collaboration is needed among research
groups to further stimulate LBD research. LBD is still an important research
theme; deep learning for improving results of LBD systems could be a
scientific frontier in the next few years.

\section*{Acknowledgements}
The authors thank Petra Hrovat Hristovski for proof-reading the
manuscript. The authors also thank Halil Kilicoglu for his helpful comments
and suggestions.

\section*{Declarations}

\subsection*{Funding}
Authors were supported by the Slovenian Research Agency (Grant No. Z5-9352 (AK) and J5-1780 (DH)).

\subsection*{Conflicts of interest}
The authors declare that they have no conflict of interest.

\subsection*{Availability of data and material}
The data set discussed in this paper has been deposited in the public
repository Zenodo (\url{https://doi.org/10.5281/zenodo.3884422}) and is freely
available to the research community.

\subsection*{Code availability}
\textsf{R} code to replicate the results of the study is accessible on the author's
GitHub page (\url{https://github.com/akastrin/lbd-review}).

\subsection*{Authors' contributions}
AK conceived the study, collected the data, performed data analysis, and wrote
the manuscript. DH contributed with critical revisions of the manuscript. Both
authors read and approved the final version of the manuscript.

\bibliographystyle{spbasic}      % basic style, author-year citations
\bibliography{ms.bib}   % name your BibTeX data base

\end{document}